\documentclass{mn2e} 
\usepackage[dvips]{graphicx} 
\usepackage{amssymb} 
\usepackage{amsbsy} 
\usepackage{txfonts} 
\newcommand{\apj}{{ApJ}} 
\newcommand{\mnras}{{MNRAS}} 
\newcommand{\aap}{{A\&A}} 
\newcommand{\msun}{{\rm M}_{\sun}} 
\newcommand{\ledd}{L_{{\rm E}}} 
\newcommand{\medd}{{\dot{M}_{\rm E}}} 
\title[]{General-relativistic model of hot accretion flows with global 
Compton cooling} 
\author[A. Nied\'zwiecki, F.-G.\ Xie, A. A. Zdziarski] 
{Andrzej Nied\'zwiecki,$^1$\thanks{E-mail: niedzwiecki@uni.lodz.pl (AN),
fgxie@shao.ac.cn (FGX), aaz@camk.edu.pl (AAZ)} 
Fu-Guo Xie,$^{2,3}$\footnotemark[1] 
and Andrzej A.~Zdziarski$^{4}$\footnotemark[1]\\ 
$^1$Department of Astrophysics, University of \L \'od\'z, Pomorska 149/153, 
90-236 \L \'od\'z, Poland\\ 
$^2$Kavli Institute for Astronomy and Astrophysics, Peking University, 
Beijing 100871, China\\ 
$^3$Key Laboratory for Research in Galaxies and Cosmology, Shanghai
 Astronomical Observatory, Chinese Academy of Sciences,\\ 
80 Nandan Road, Shanghai 200030, China\\ 
$^{4}$ Centrum Astronomiczne im.\ M. Kopernika, Bartycka 18, 
00-716 Warszawa, Poland\\ } 

\pagerange{\pageref{firstpage}--\pageref{lastpage}} 
\pubyear{2011} 
\begin{document} 
\maketitle 
\label{firstpage} 
 
\begin{abstract} 
We present a model of optically thin, two-temperature, accretion flows using an exact Monte Carlo treatment of global Comptonization, with seed photons from synchrotron and bremsstrahlung emission, as well as with a fully general relativistic description of both the radiative and hydrodynamic processes. We consider accretion rates for which the luminosities of the flows are between $\sim 10^{-3}$ and $10^{-2}$ of the Eddington luminosity. The black hole spin parameter strongly affects the flow structure within the innermost $\simeq 10$ gravitational radii. The resulting large difference between the Coulomb heating in models with a non-rotating and a rapidly rotating black hole is, however, outweighed by a strong contribution of compression work, much less dependent on spin.  The consequent reduction of effects related to the value of the black spin is more significant at smaller accretion rates. For a non-rotating black hole, the compressive heating of electrons dominates over their Coulomb heating, and results in an approximately constant radiative efficiency of $\approx 0.4$ per cent in the considered range of luminosities. For a rapidly rotating black hole, the Coulomb heating dominates, the radiative efficiency is $\simeq 1$ per cent and it slightly increases (but less significantly than estimated in some previous works) with increasing accretion rate. 
Our study neglects the direct heating of electrons, which effect can lead to larger differences between the radiative 
properties of models with a non-rotating and a rapidly rotating black hole than estimated here.
Flows with the considered parameters produce rather hard spectra, with the photon spectral index $\Gamma \sim 1.6$, and with high energy cut-offs at several hundred keV. We find an agreement between our model, in which the synchrotron emission is the main source of seed photons, and observations of black-hole binaries in their hard states and AGNs at low luminosities. In particular, our model predicts a hardening of the X-ray spectrum with increasing luminosity, as indeed observed below $\sim 0.01 \ledd$ or so in both black-hole binaries and AGNs. Also, our model approximately reproduces the luminosity and the slope of the X-ray emission in Cen A. 

\end{abstract} 
\begin{keywords} 
accretion, accretion discs -- black hole physics -- X-rays: binaries
-- X-rays: general. 
\end{keywords} 
 
\section{Introduction} 
\label{intro} 

Optically thin, two-temperature accretion flows have been
considered as an explanation of a variety of black hole systems 
and a substantial work has been done for investigation of their
dynamical and spectral properties (see, e.g., Narayan
\& McClintock 2008 for a review). However, the developed models still involve
several approximations which significantly reduce their
accuracy. These involve the use of a pseudo-Newtonian potential of Paczy\'nski \& Wiita (1980), which fails in the innermost region
(particularly if rotation of the black hole is considered),
where most of the gravitational energy is dissipated, as well as 
local approximations of Comptonization, which appear to be particularly
incorrect in optically thin flows (see Xie et al.\ 2010; hereafter
X10). Previous attempts to improve these two weaknesses 
are discussed in Section \ref{previous}.

In this paper, we extend our previous treatment of global Comptonization from X10 by using the hydrodynamical model from Manmoto (2000; hereafter M00), and develop a self-consistent model involving a fully general-relativistic (GR) description of both the hydrodynamical and the radiative processes. We consider moderate values of accretion rate, at which Coulomb coupling between ions
and electrons is relatively weak compared to the viscous heating of
ions and, therefore, the ion temperature (and hence the total
pressure) is not affected significantly by the details of the
description of radiative processes. This allows for a slightly
simplified treatment of the flow structure in our computations 
leading to a self-consistent solution. 

Already the foundational papers proposed that tenuous, two-temperature flows
may be responsible for the hard spectral states of black-hole
binaries (Ichimaru 1977; see also Narayan \& Yi 1995) and for the low nuclear luminosities in
radio galaxies with large radio lobes (Rees et al.\ 1982). We
consider two ranges of the key parameters (black hole mass and accretion
rate) which should be relevant for these two major application
areas. For each of these two cases, we illustrate the impact of the
black hole spin by considering a non-rotating and a maximally-rotating
black hole. On the other hand, we neglect here the dependence on some other
parameters of hot-flow models. 

In particular, we consider only weakly magnetized flows, with the magnetic pressure of 1/10th of the total pressure,
and we neglect the direct viscous heating of electrons; see Section \ref{delta} for a discussion of these two assumptions.
Furthermore, we neglect outflows, which may play an important role in some systems (e.g.\ Yuan, Quataert \& Narayan 2003). Then, our solutions are strongly dominated by the advection of energy by ions, as in the original formulation of the advection-dominated accretion flow (ADAF) model by, e.g., Narayan \& Yi (1995).

In this paper, we focus on modelling an innermost part of an accretion
flow, namely inside $10^3 R_{\rm g}$, where $R_{\rm g} = GM/c^2$ is the gravitational radius and $M$ is the black-hole mass, where the bulk of the observed radiation is produced. Global Comptonization has a net effect of Compton heating outside $\sim\! 10^4 R_{\rm g}$ (e.g.\ Park \& Ostriker 2001, Yuan, Xie \& Ostriker 2009, Yuan \& Li 2011), which effect is not considered here.

\section{The model} 
\label{model} 

We consider a black hole, characterised by its mass, $M$, and
angular momentum, $J$, surrounded by a geometrically thick
accretion flow with an accretion rate, $\dot M$. We define the following
dimensionless parameters: $r = R / R_{\rm g}$, $a = J / (c R_{\rm g} M)$, 
$\dot m = \dot M / \dot \medd$, where $\dot \medd= \ledd/c^2$ and $\ledd \equiv 4\pi GM m_{\rm p} c/\sigma_{\rm T}$ 
is the Eddington luminosity. The inclination angle of the line
of sight to the symmetry axis is given by $\mu_{\rm obs} \equiv \cos
\theta_{\rm obs}$. 

We refine here our previous study presented in
X10 by including a fully GR hydrodynamical description of the
flow. Our GR hydrodynamical model follows strictly M00,
except for a small difference in the advective term, as noted below
equation (\ref{eq:qint}). We have thoroughly tested our hydrodynamical model 
by comparing it with the numerical code and results of Li et al.\ (2009),
which work also follows M00. Other differences with respect to X10 involve: 

\noindent (i) the vertical structure of the flow: we assume here
that the temperature and velocity components are vertically uniform,
while the density distribution is given by $\rho(R,z)=\rho(R,0)
\exp(-z^2/2H^2)$, where $H$ is the scale height at $r$; in X10 we
assumed a tangentially uniform structure of the flow; 

\noindent (ii) the direct viscous heating of electrons, which is
neglected here and included in X10; 

\noindent (iii) an outflow, neglected here and included (with a
large magnitude) in X10. 

We consider the pairs of the main parameters of 
($M =2 \times 10^8 \,\msun$, $\dot m=0.1$) and ($M=10 \,\msun$, $\dot m=0.5$), 
which correspond to low-luminosity AGNs and hard states of black-hole binaries, respectively. For each pair of ($M$, $\dot m$) we consider two values of the spin parameter, $a=0$ and $a=0.998$. We assume the viscosity parameter of $\alpha=0.3$ and the ratio of the gas pressure 
(electron and ion) to the total pressure of $\beta_{\rm B} = 0.9$.
The latter parameter determines the strength of the magnetic field in
the accretion flow, and the synchrotron emission can then be
determined. For it, we follow the method of Narayan \& Yi (1995),
which was applied also by M00 and X10. 

We define the compressive heating rate of electrons per unit area (in the local reference frame) of the flow,
\begin{equation} 
\Lambda_{\rm compr} = - {\dot M p_{\rm e} \over 2 \pi R \rho} 
{{\rm d}\ln \rho \over {\rm d}R},
\label{eq:compr}
\end{equation}
and the rate (per unit area) of advection of the internal energy of
electrons,
\begin{equation}
Q_{\rm int} = - {\dot M p_{\rm e} \over 2 \pi R \rho (\Gamma_{\rm e} - 1)} 
{{\rm d}\ln T_{\rm e} \over {\rm d} R}, 
\label{eq:qint}
\end{equation}
where $\rho$ is the mass density and $p_{\rm e}$ is the electron
pressure in the midplane, $\Gamma_{\rm e}$ is the effective
adiabatic index [see, e.g., eq.\ (62) in M00 for the
definition]; for simplicity of notation we hereafter skip the 'e' subscripts in symbols denoting electron heating/cooling rates. Note that in our definition $\dot M > 0$. Then, $\Lambda_{\rm compr}$ is always positive (therefore, we define it as
a heating term) as the density gradient is always negative in our
models. In all models, $Q_{\rm int}$ is negative (representing the
release of internal energy) in the innermost region, at larger
distances it is positive (the internal energy is stored) or vanishes.
Following the usual notation, we define also the electron advection rate,
$Q_{\rm adv} \equiv Q_{\rm int} - \Lambda_{\rm compr}$, which is
typically negative (i.e.\ the advective effects result in effective
heating) except for an outer (beyond $r_{\rm s}$, see below) region
in the local models. The above forms of $\Lambda_{\rm compr}$ and
$Q_{\rm int}$ follow directly from eqs.\ (54) and (60) in M00. 
We do not use the simplifying approximation, ${\rm d}\ln H /
{\rm d}\ln R = 1$, adopted in M00; we checked that it
yields deviations of the advective term from the actual values by a factor of a few.

With the above definitions, the electron energy equation is
\begin{equation} 
0 = \Lambda_{\rm ie} + \Lambda_{\rm compr} - Q_{\rm rad} - Q_{\rm int},
\label{eq:energy} 
\end{equation}
where $Q_{\rm rad}$ is the radiative cooling rate per unit area
(including synchrotron, $Q_{\rm synch}$, and bremsstrahlung,
$Q_{\rm brem}$, and their Comptonization, $Q_{\rm Compt}$) and
$\Lambda_{\rm ie}$ is the electron heating rate per unit area by ions
via Coulomb collisions.

The values of $\Lambda_{\rm ie}$, $\Lambda_{\rm compr}$ and $Q_{\rm rad}$ 
integrated over the whole body of the flow are denoted by
$\Lambda_{\rm ie,tot}$, $\Lambda_{\rm compr,tot}$ and $Q_{\rm rad,tot}$, respectively. Due to the effect of capturing by the
black hole and the gravitational redshift (see Section \ref{collimation} for details), the total power radiated by the flow, $Q_{\rm rad,tot}$, is higher than the luminosity, $L$, detected far away from the flow.  

For completeness we give also the ion energy equation: 
\begin{equation} 
0 = Q_{\rm diss}  + \Lambda_{\rm compr,i} - \Lambda_{\rm ie} - Q_{\rm int,i},
\end{equation}
where the compressive heating and the advection of the internal energy of
ions, respectively, are given by
\begin{equation} 
\Lambda_{\rm compr,i} = - {\dot M p_{\rm i} \over 2 \pi R \rho} 
{{\rm d}\ln \rho \over {\rm d}R},~~~
Q_{\rm int,i} = - {\dot M p_{\rm i} \over 2 \pi R \rho (\Gamma_{\rm i} - 1)} 
{{\rm d}\ln T_{\rm i} \over {\rm d} R}, 
\end{equation}
and the viscous dissipation rate, per unit area, is given by  
\begin{equation} 
Q_{\rm diss} = - \frac{\gamma_\phi^4A^2}{r^7}(2 \upi)^{1/2} \alpha p  H \frac{{\rm d}\Omega}{{\rm d}r}, 
\label{eq:qdiss} 
\end{equation}
where $p =  (p_{\rm i} + p_{\rm e})/\beta_{\rm B}$, $p_{\rm i}$ is the ion pressure,
$\Gamma_{\rm i}$ is the ion adiabatic index, $\Omega$ is the angular velocity of the flow, $\gamma_\phi$ is the Lorentz factor of the azimuthal motion and $A=r^4 + r^2a^2 + 2ra$. The above form of $Q_{\rm diss}$ results from the usual assumption that the viscous stress is proportional to the total pressure,
with the proportionality coefficient $\alpha$; cf.\ eqs.\ (48) and (54) in M00. This is almost certainly an oversimplified approach; see Gammie \& Popham (1998) for a discussion of various issues related to implementing
a proper description of the viscous stress in GR models. The magneto-rotational
instability (MRI) is now widely believed to be an origin of the  
viscous stress and numerical simulations of this process (e.g.\ Krolik, Hawley
\& Hirose 2005) show radial
profiles of the stress which are different from analytic prescriptions considered in the accretion theory.  
The uncertainty regarding the form of the stress is the major shortcoming
in attempts to develop an analytical description of accretion flows. 
The form of the viscous stress, $\propto \alpha p$, is assumed here for computational
simplicity. It has an unphysical property of leading to a negative dissipation rate close
to the event horizon in some cases.   
However, this particular weakness of our model appears to be not 
important for our results. Namely, we find that $Q_{\rm diss} < 0$ occurs at $r \la 3$ only in our models  with $a=0$; in these models the heating of ions is strongly dominated by the compression work at $r<10$. Therefore, details of the viscous heating of ions do not affect the flow structure.
Clearly, the accuracy of $Q_{\rm diss}$ would be much more important in models taking into account direct viscous heating of electrons.

Our modelling of Comptonization makes use of a Monte Carlo (MC) method taking into account all effects relevant for the Kerr metric (Nied\'zwiecki 2005;
see also Nied\'zwiecki \& Zdziarski 2006 and X10). The seed
photons are generated from synchrotron and bremsstrahlung radial
emissivities of the flow, taking into account the vertical profile of 
the density; their transfer and energy gains in consecutive
scatterings are affected by both the special relativistic and
gravitational effects. 

\begin{figure*} 
\centerline{\hspace{-0.05cm} 
 \includegraphics[width=5.8cm]{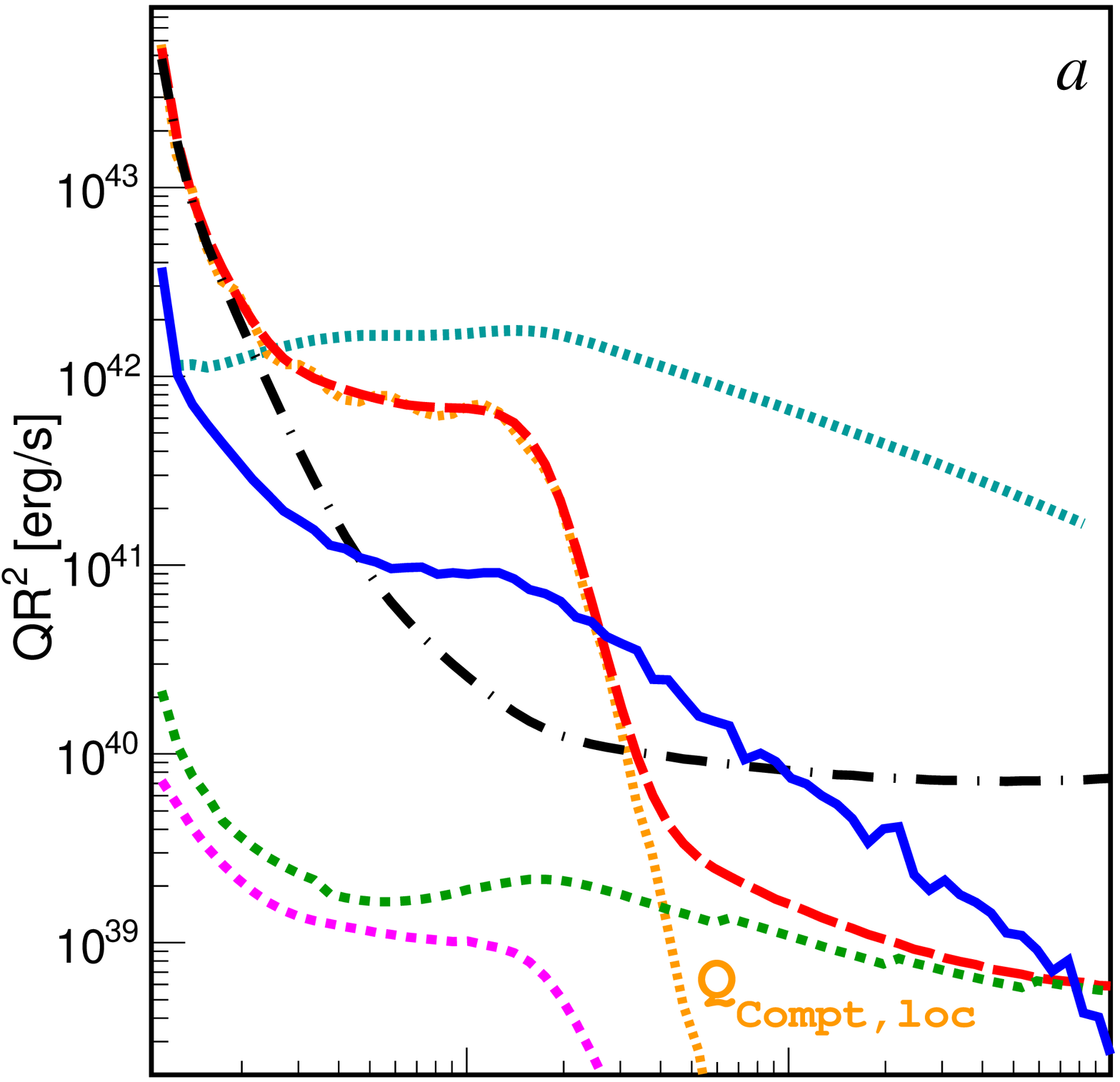}\hspace{0.001cm} 
 \includegraphics[width=5.8cm]{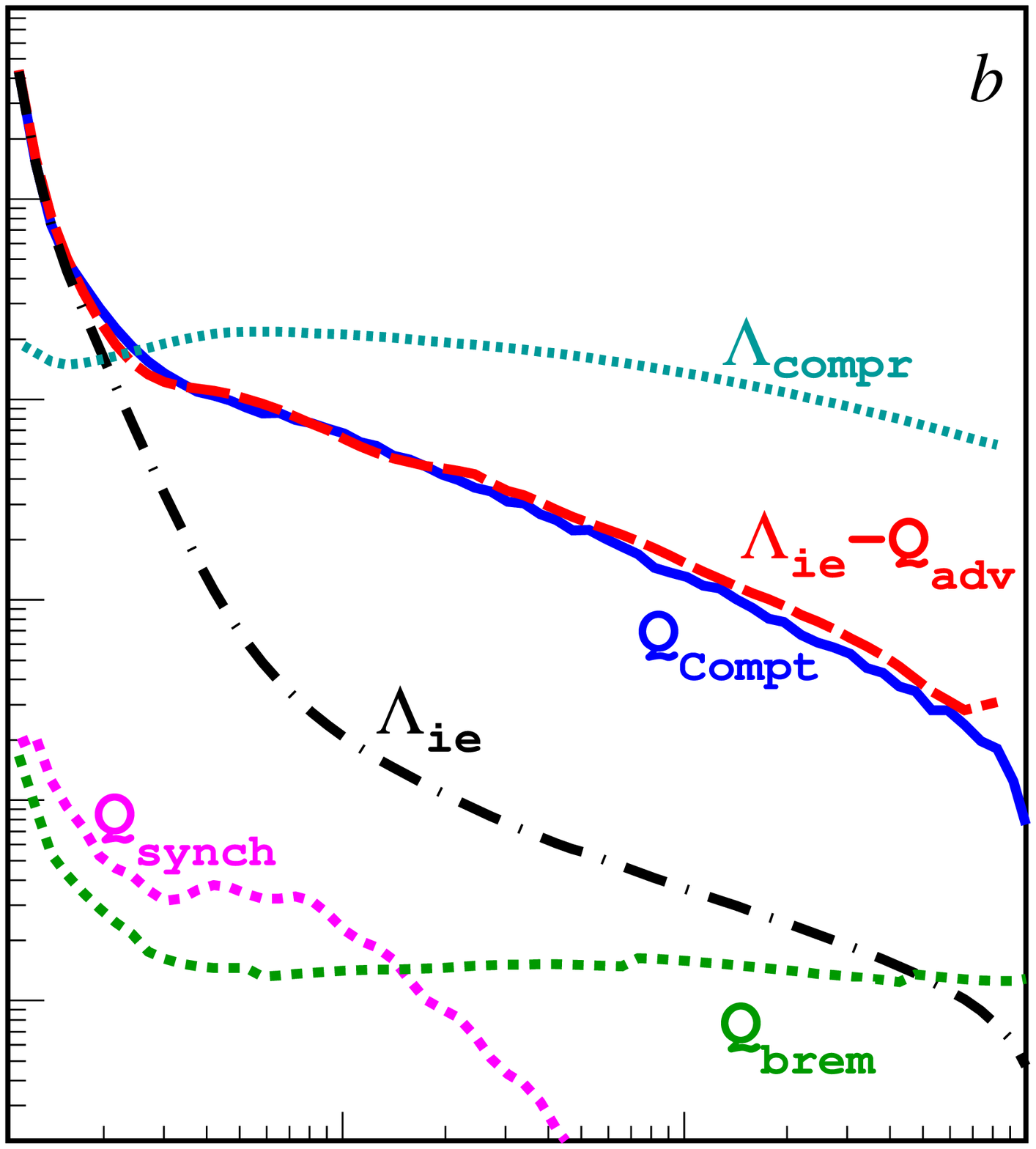}\hspace{-0.15cm} 
 \includegraphics[width=5.8cm]{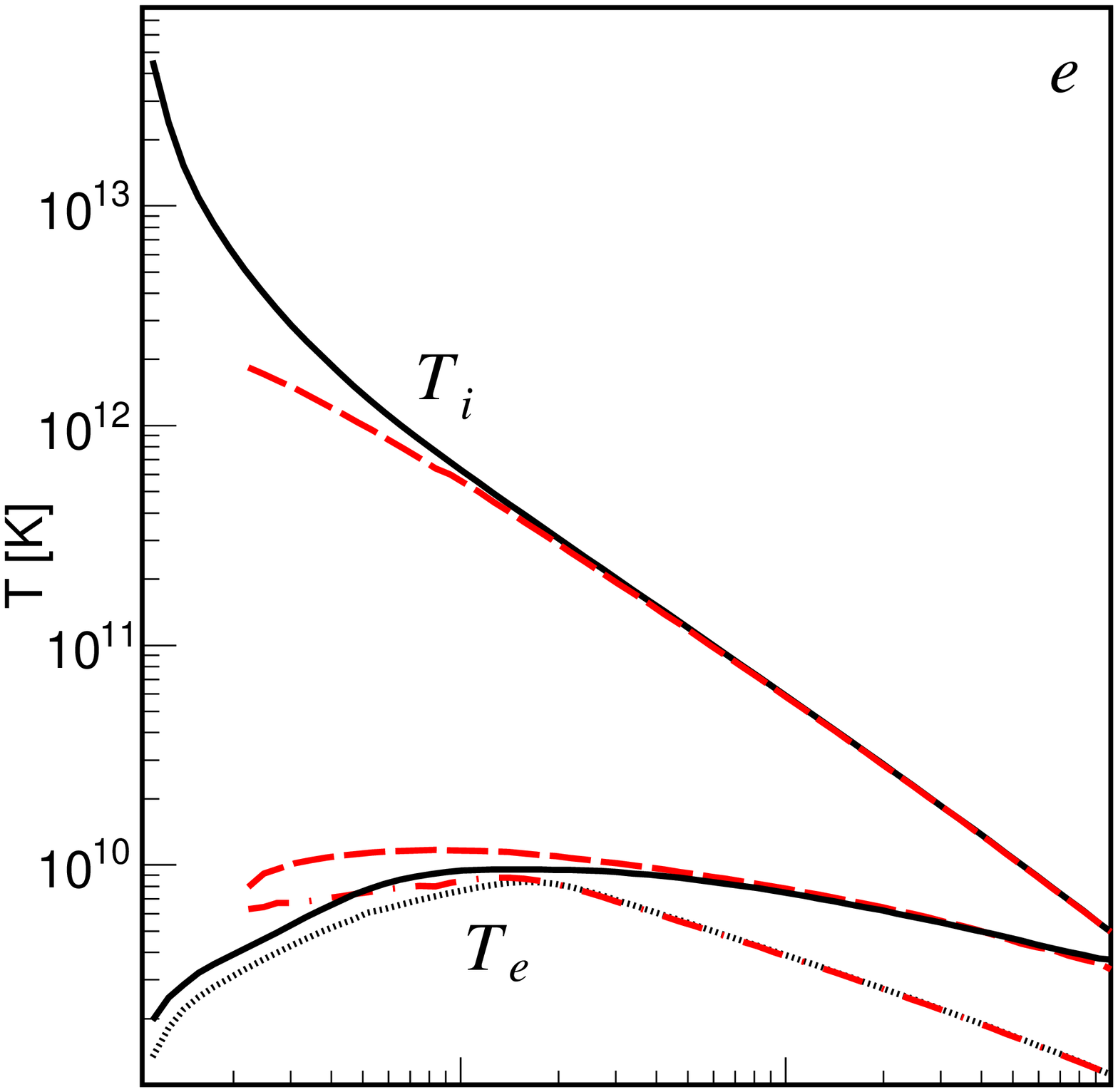}} 
\centerline{\includegraphics[width=5.8cm]{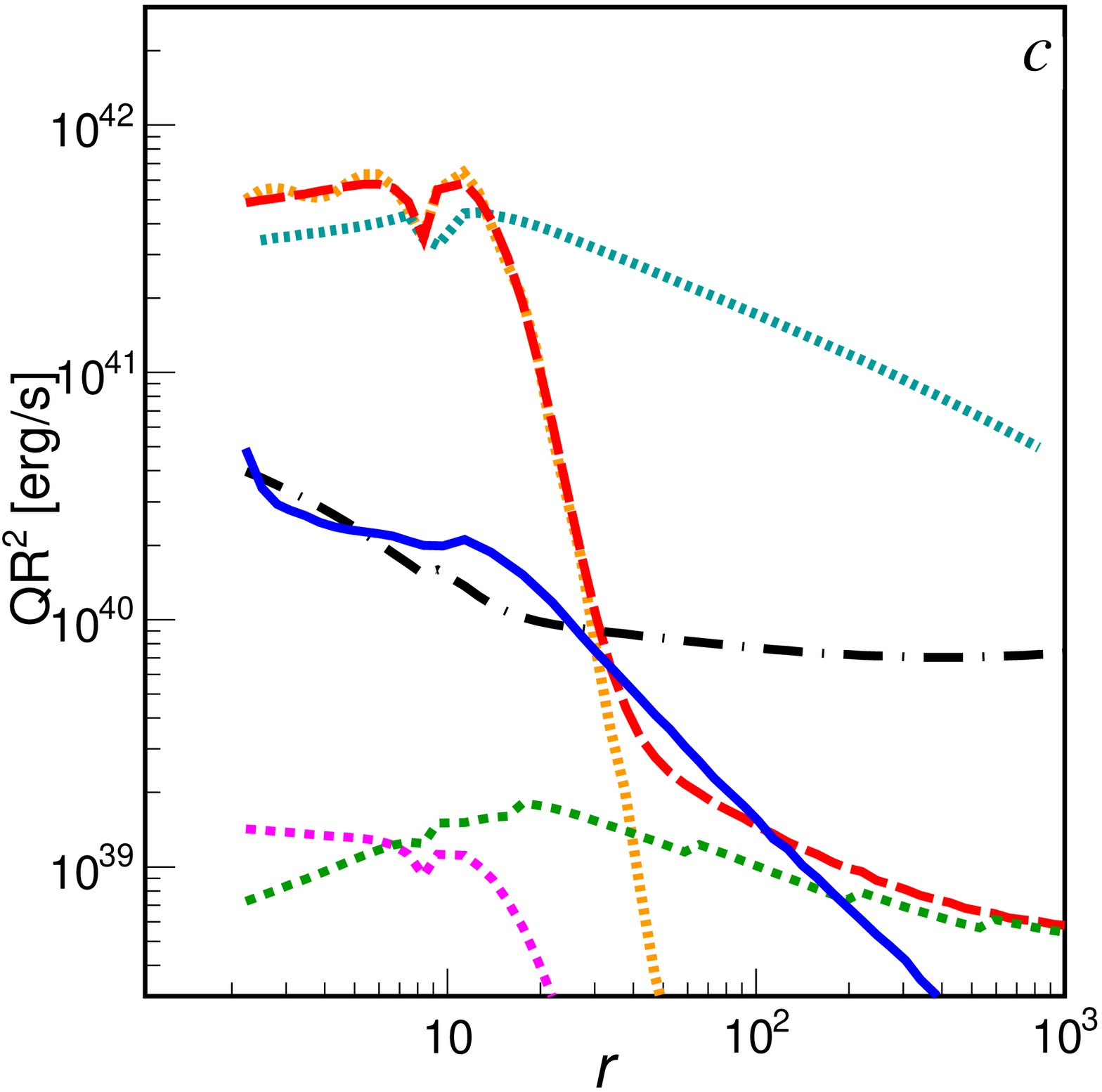}\hspace{0.001cm} 
 \includegraphics[width=5.8cm]{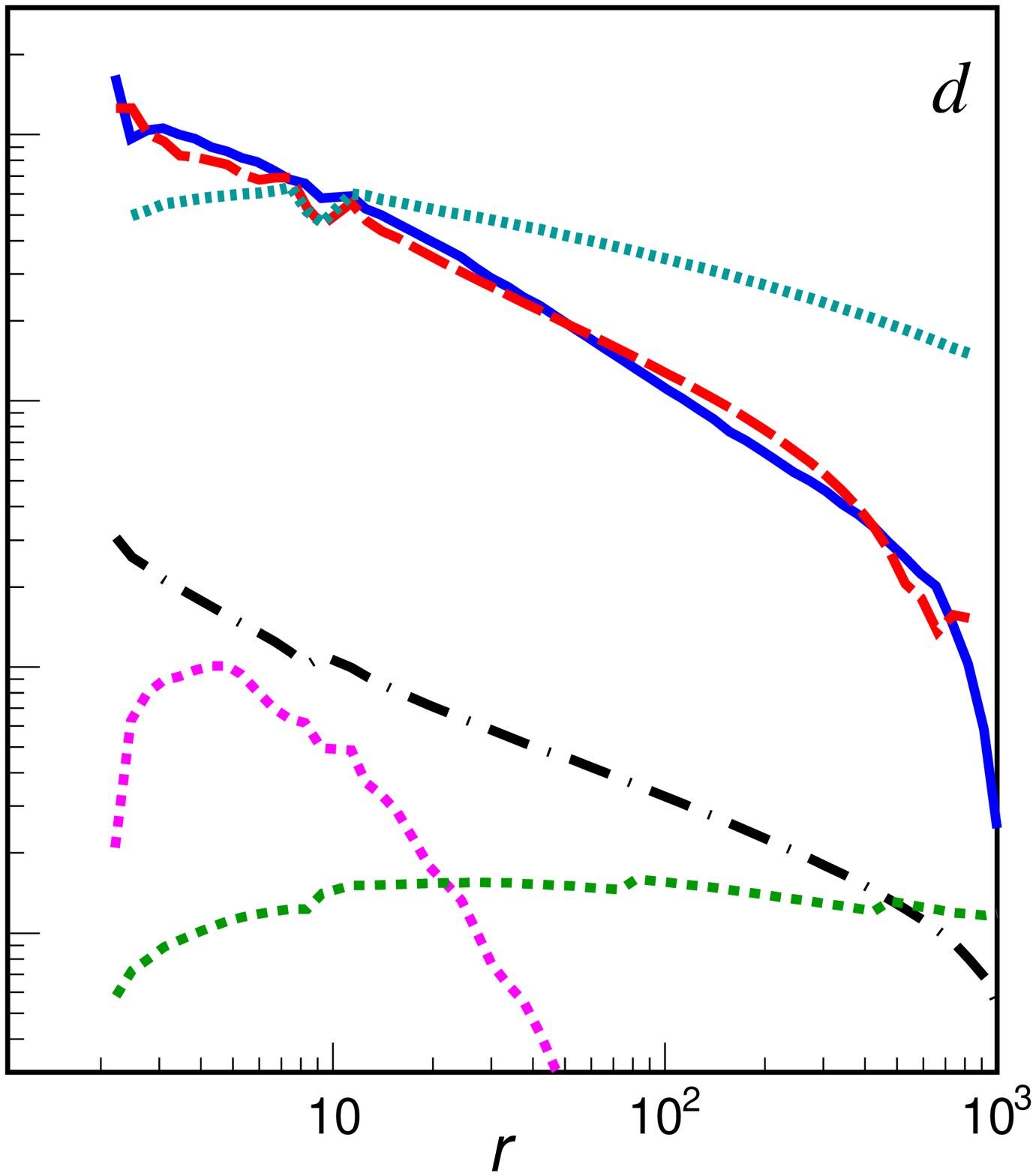}\hspace{-0.15cm} 
 \includegraphics[width=5.8cm]{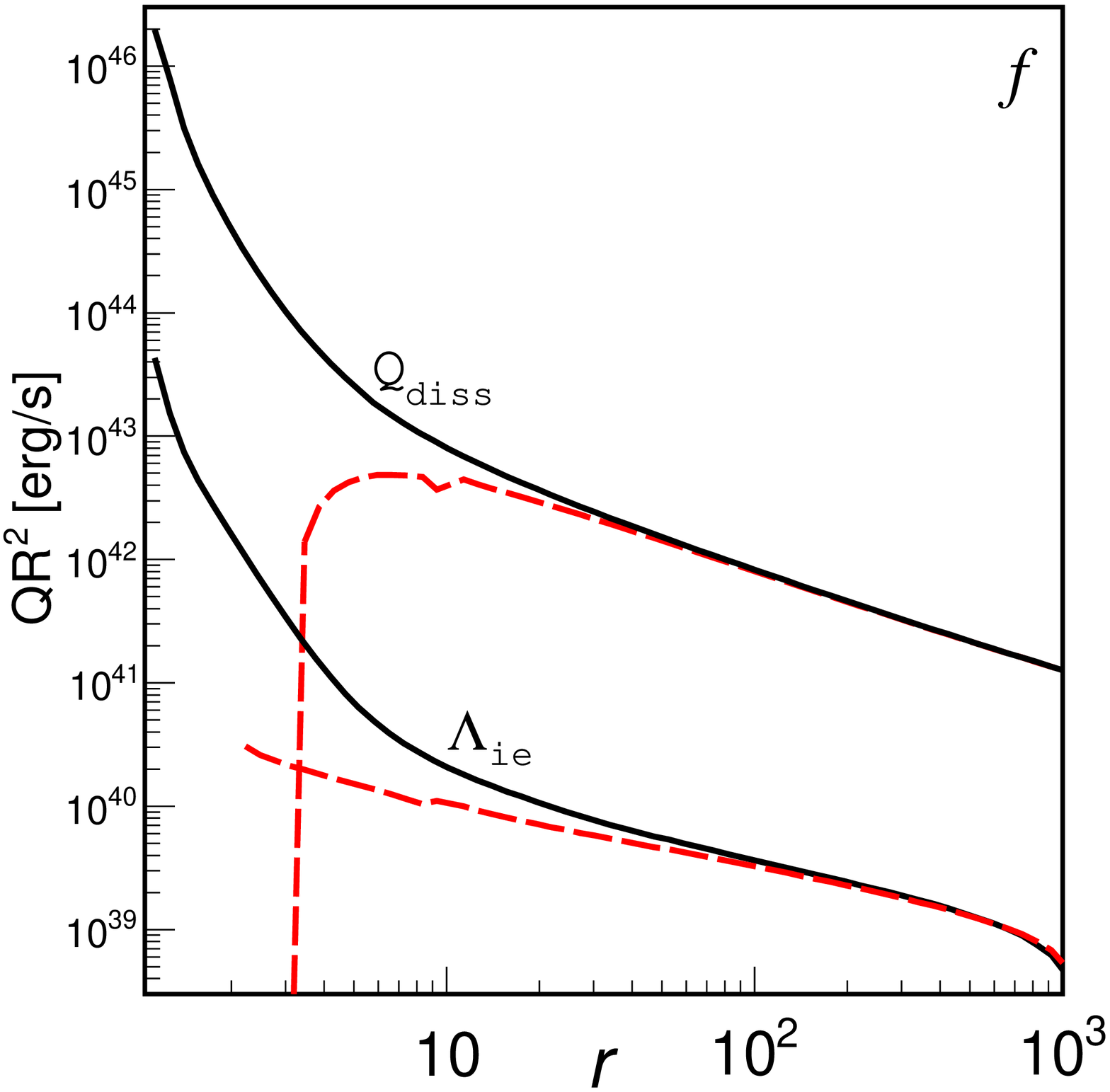}} 
\caption{Comparison between initial and final solutions in models 
 with $M =2 \times 10^8 \!\msun $ and $\dot m=0.1$. Panels (a--d)
 show the electron heating and cooling rates for the initial
 (a,c) and final (b,d) solutions. $Q$ denotes a vertically 
 integrated rate, so $QR^2$ gives cooling/heating rates (per unit
 volume) times volume. The upper (a,b) and lower (c,d) panels are 
 for $a=0.998$ and $a=0$, respectively. The short dashed curves 
 show the synchrotron (magenta online) and bremsstrahlung (green)
 emissivities. The dot-dashed (black) curves show the Coulomb
 rate, $\Lambda_{\rm ie}$. The long dashed (red) curves show the
 effective heating rate of electrons, $\Lambda_{\rm ie} - Q_{\rm
 adv}$. The dotted (cyan) curves show the compressive heating
 rate, $\Lambda_{\rm compr}$. The solid (blue) curves show the
 global Compton cooling rate, $Q_{\rm Compt}$. The dotted (orange)
 curves in panels (a,c) show the Compton cooling rate in the local
 slab approximation, $Q_{\rm Compt,loc}$. Panel (e) shows the ion
 (upper curves) and electron (lower curves) temperatures. The solid
 (black) and dashed (red) curves correspond to our final,
 self-consistent solution for $a=0.998$ and $a=0$, respectively
 ($T_{\rm i}$ shown here is for both the initial and final
 solution); the dotted (black, $a=0.998$) and dot-dashed (red,
 $a=0$) curves show $T_{\rm e}$ for the initial solutions. Panel
  (f) shows the dissipative heating rate of ions (upper curves) and  
  the Coulomb rate between ions and electrons (lower curves) for
  $a=0.998$ (solid, black curves)  and $a=0$ (dashed, red). 
} 
\label{fig:rates} 
\end{figure*}

The MC Comptonization simulation and the solution of structure equations are treated separately in our model. Therefore, we have to iterate between their results to find a mutually consistent solution. In principle, a full set of
hydrodynamical equations should be solved after each MC simulation, 
as in X10. However, in the current study we apply a slightly 
simplified procedure, motivated by the fact that the structure of
the flow is fully determined by the ion energy equation (this
condition is, however, satisfied only within inner several
hundred $R_{\rm g}$, see below), while the electron energy equation
determines the electron temperature. Under such conditions, our
procedure of refining the description of Compton cooling should
affect only the electron temperature and not the other flow
parameters ($T_{\rm i}$, $\rho$, $H$, and the velocity field). Then, we first solve the dynamical structure with a local approximation of Compton cooling and then we iterate between the MC Comptonization results and the solutions of
the electron energy equation, keeping the flow parameters, except
for $T_{\rm e}$, unchanged. After each MC simulation, we determine 
the radially-dependent Compton cooling. We then use this rate to solve
the energy equation (\ref{eq:energy}), taking into account the
dependence of all relevant processes (synchrotron and bremsstrahlung
emissivities, Coulomb exchange rate, advective terms) on $T_{\rm e}$. 
We repeat the above procedure until it converges. 

For the initial solution, the outer boundary of the flow is set at $r
= 2 \times 10^4$ (see section 2.4 in M00 for details). The
MC simulations are performed within $r_{\rm out}=1000$, using the hydrodynamical solution at $r \le r_{\rm out}$; note that details of this inner part of the
solution should not depend on the outer boundary condition (see
e.g.\ fig.\ 5 in Narayan, Kato \& Honma 1997). For our final
solution, we balance the electron energy equation up to $r \simeq
700$. Due to boundary effects in MC simulation, combined
with strong sensitivity of the advective term on the temperature
gradient, the solution is very unstable at $r$ close to $r_{\rm out}$. 
Furthermore, we note that an assumption underlying our computational
procedure fails in these outer regions. Namely, all our final solutions
are characterised by an electron temperature larger (typically by a
factor of $\sim 3$ at $r \la r_{\rm out}$) than the initial solutions
and the electron pressure can give a non-negligible contribution
beyond several hundred $R_{\rm g}$. In that region, the
scale height would increase, which effect is not taken into account
here; however, we expect it to be a rather minor, $\la 10$ per cent, effect --
such a change of $H/R$ was found, in response to a similar in
magnitude (but opposite) change of $T_{\rm e}$, in X10. For one set
of parameters, namely $M=10\, \msun$, $\dot m = 0.5$ and $a=0$, 
$T_{\rm e}$ approaches $T_{\rm i}$ at $r \simeq 1000$, so the flow loses a
two-temperature structure. We skip here more careful investigation
of the outer region because it gives a negligible contribution to
the observed radiation. Furthermore, a detailed study of the outer
region should involve a proper description of additional, poorly
understood effects (like a mechanism of transition from an
optically thick disc to a tenuous flow; e.g., in hard states such a
transition is likely to occur at $r<1000$) which are beyond the
scope of this paper.  

\begin{figure*} 
\centerline{\hspace{-0.05cm} 
 \includegraphics[width=5.8cm]{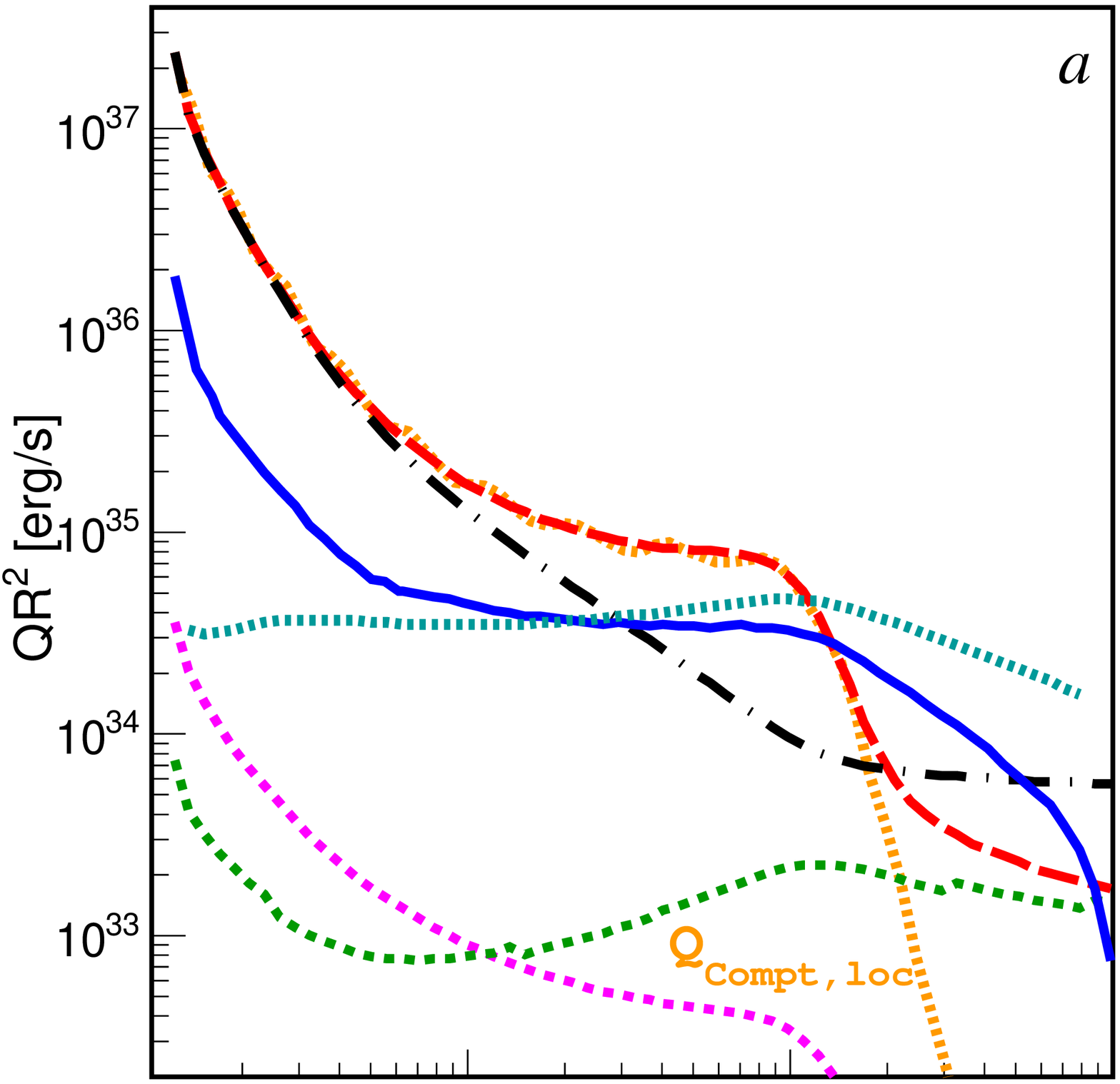}\hspace{0.001cm} 
 \includegraphics[width=5.8cm]{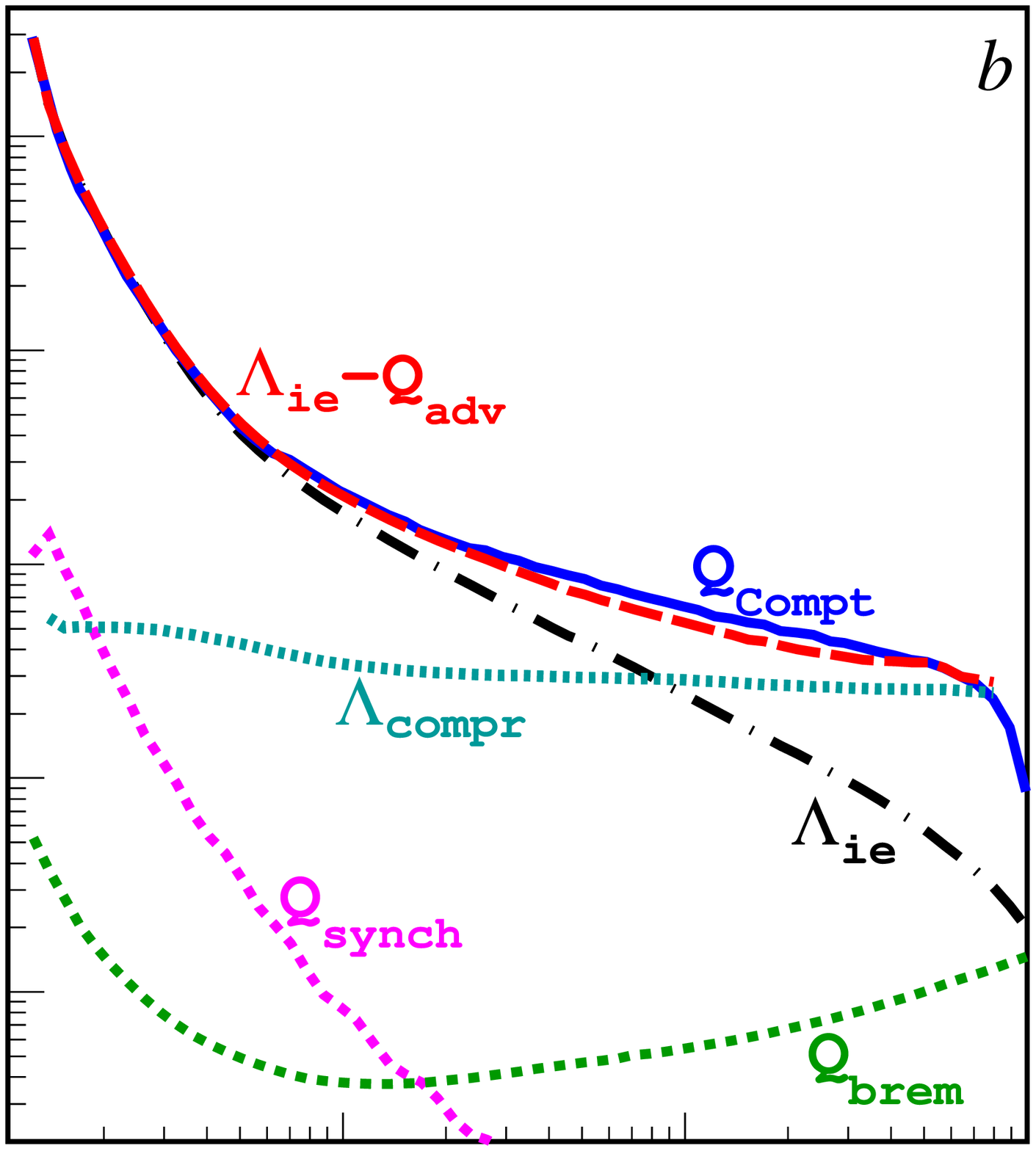}\hspace{-0.15cm} 
 \includegraphics[width=5.8cm]{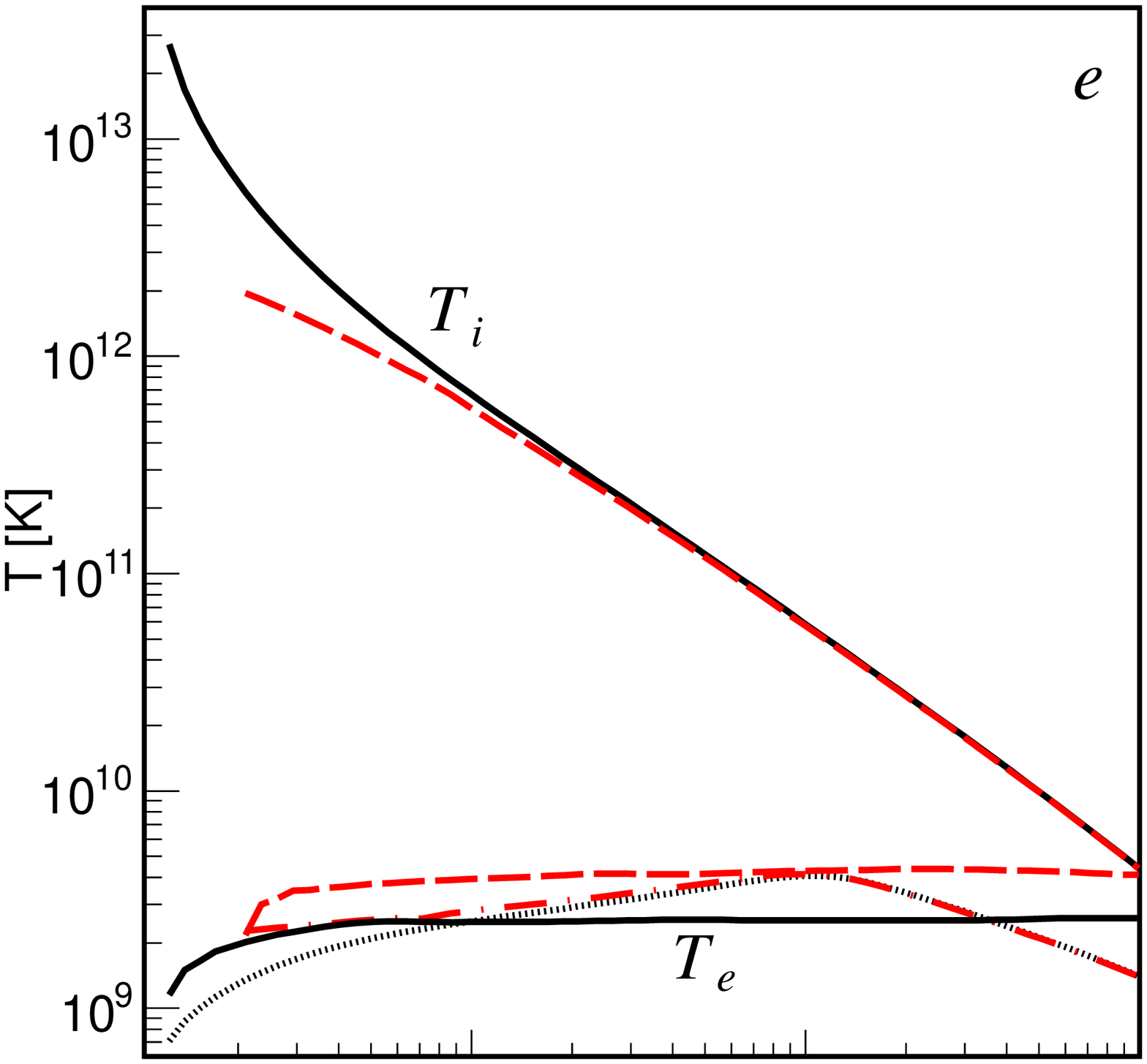}} 
\centerline{\includegraphics[width=5.8cm]{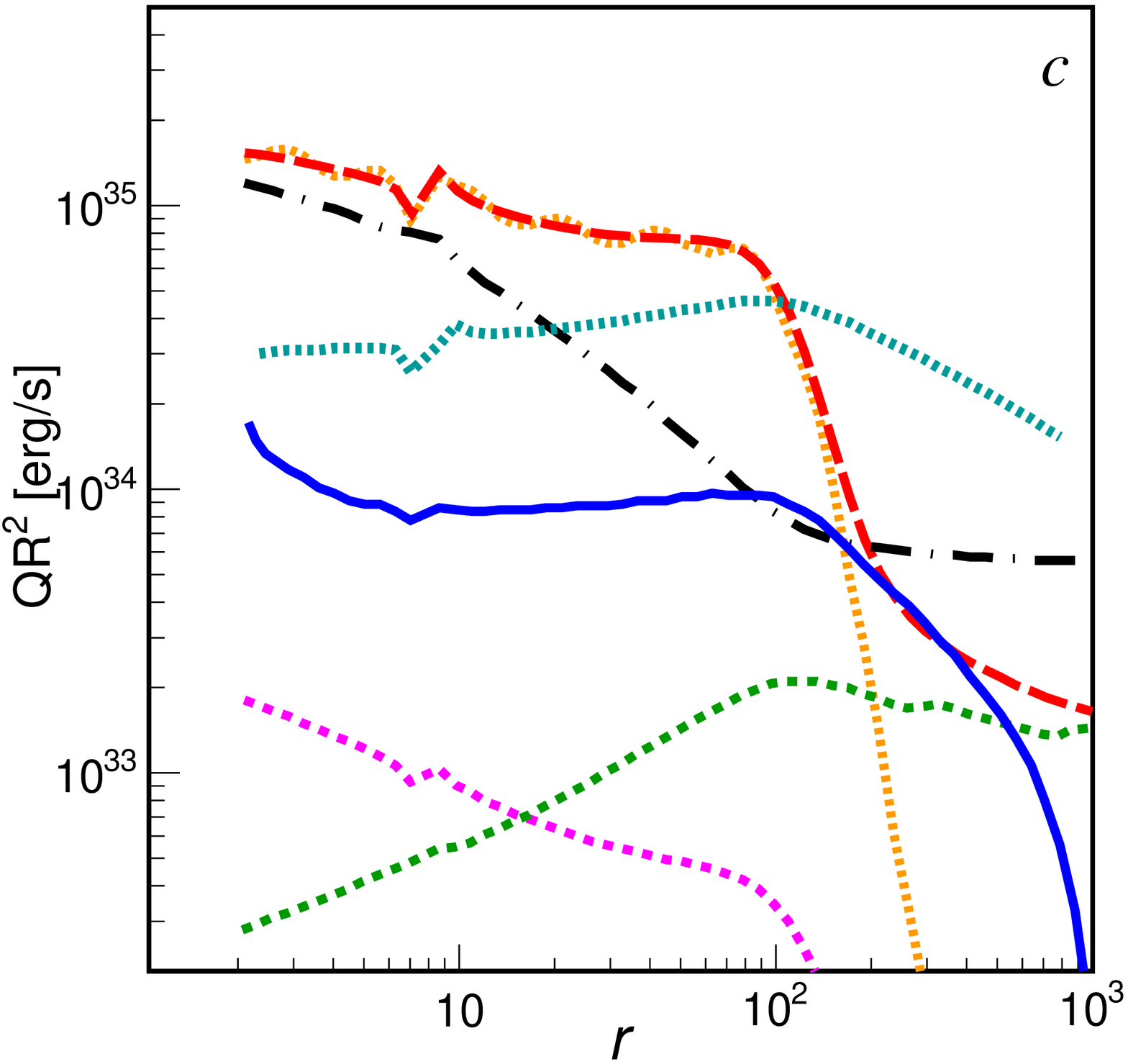}\hspace{0.001cm} 
 \includegraphics[width=5.8cm]{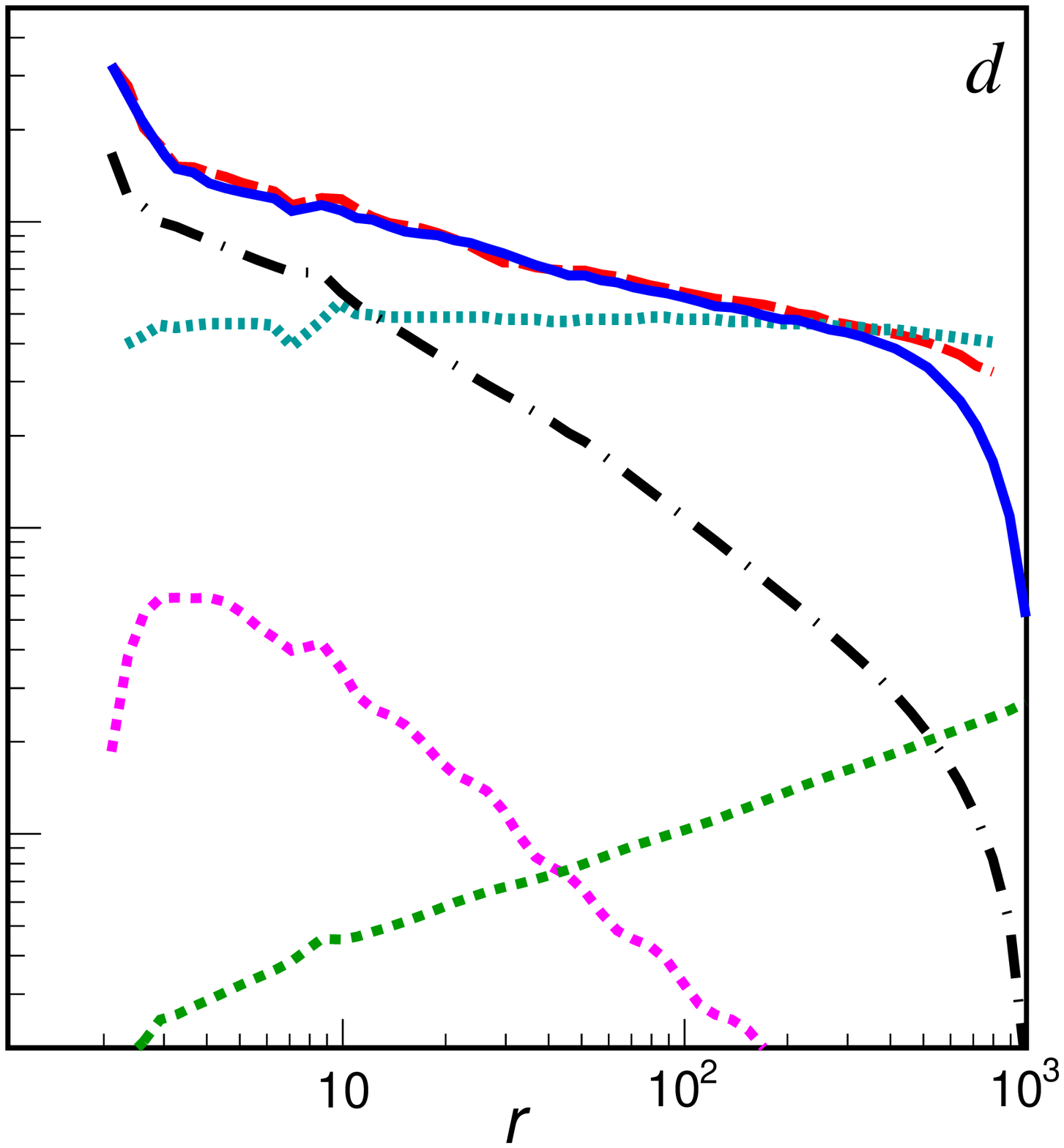}\hspace{-0.15cm} 
 \includegraphics[width=5.8cm]{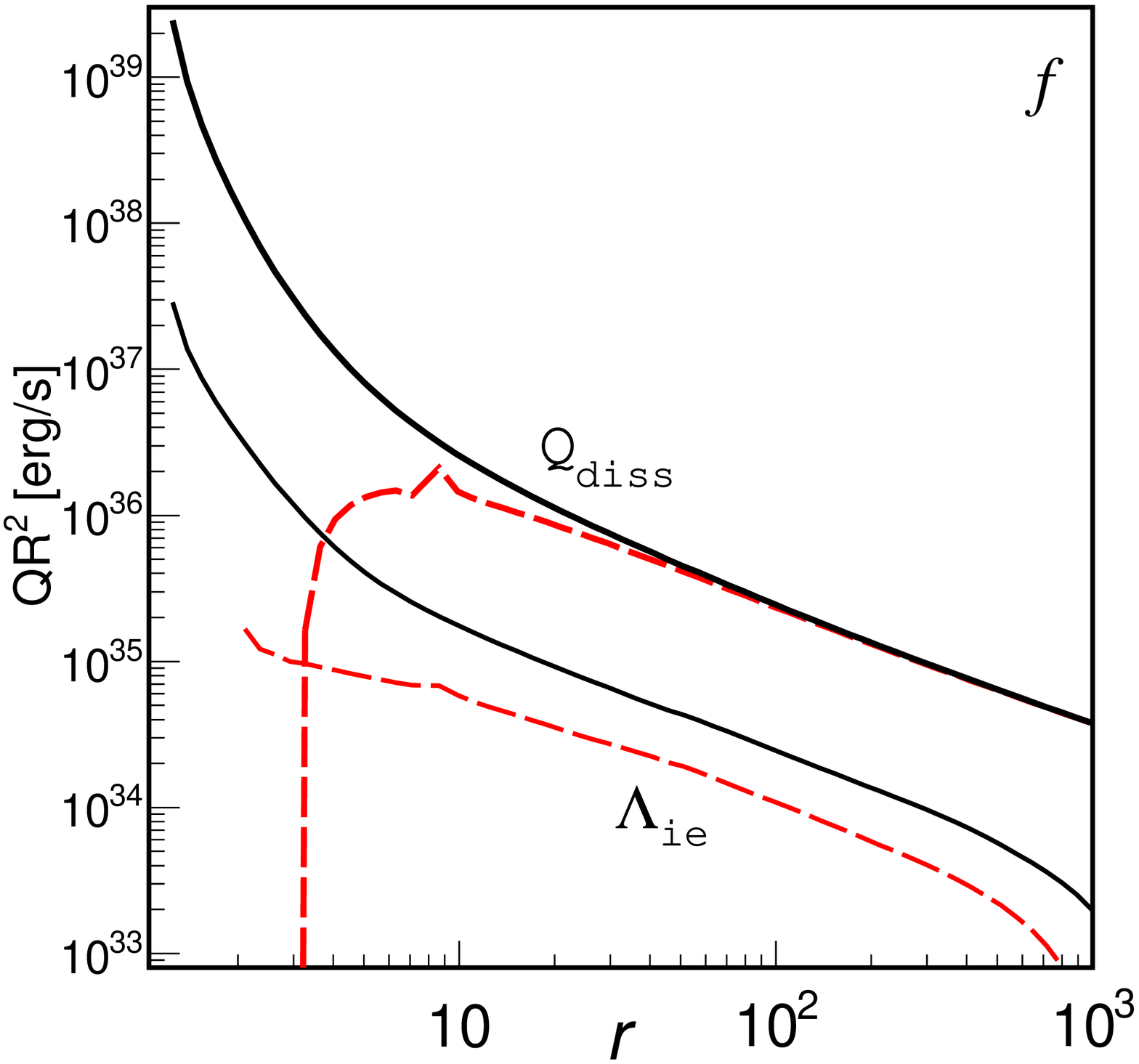}} 
\caption{Same as in Fig.\ 1 but for $M = 10 \,\msun$ and $\dot
 m=0.5$.} 
\label{fig:rates_stel} 
\end{figure*}

\section{Results}

\begin{figure*} 
\centerline{\includegraphics[width=4.2cm]{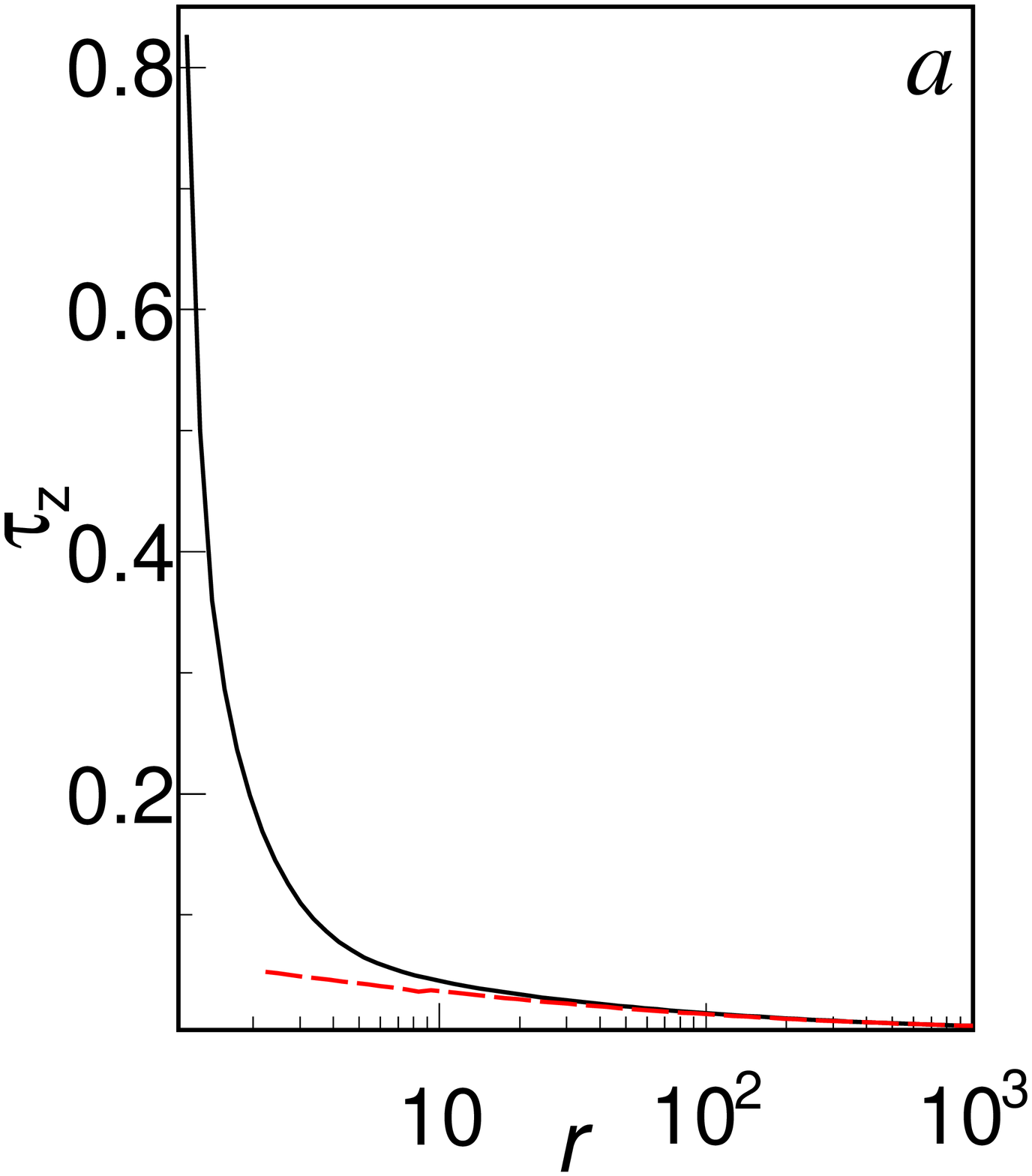}\hspace{0.12cm} 
 \includegraphics[width=4.2cm]{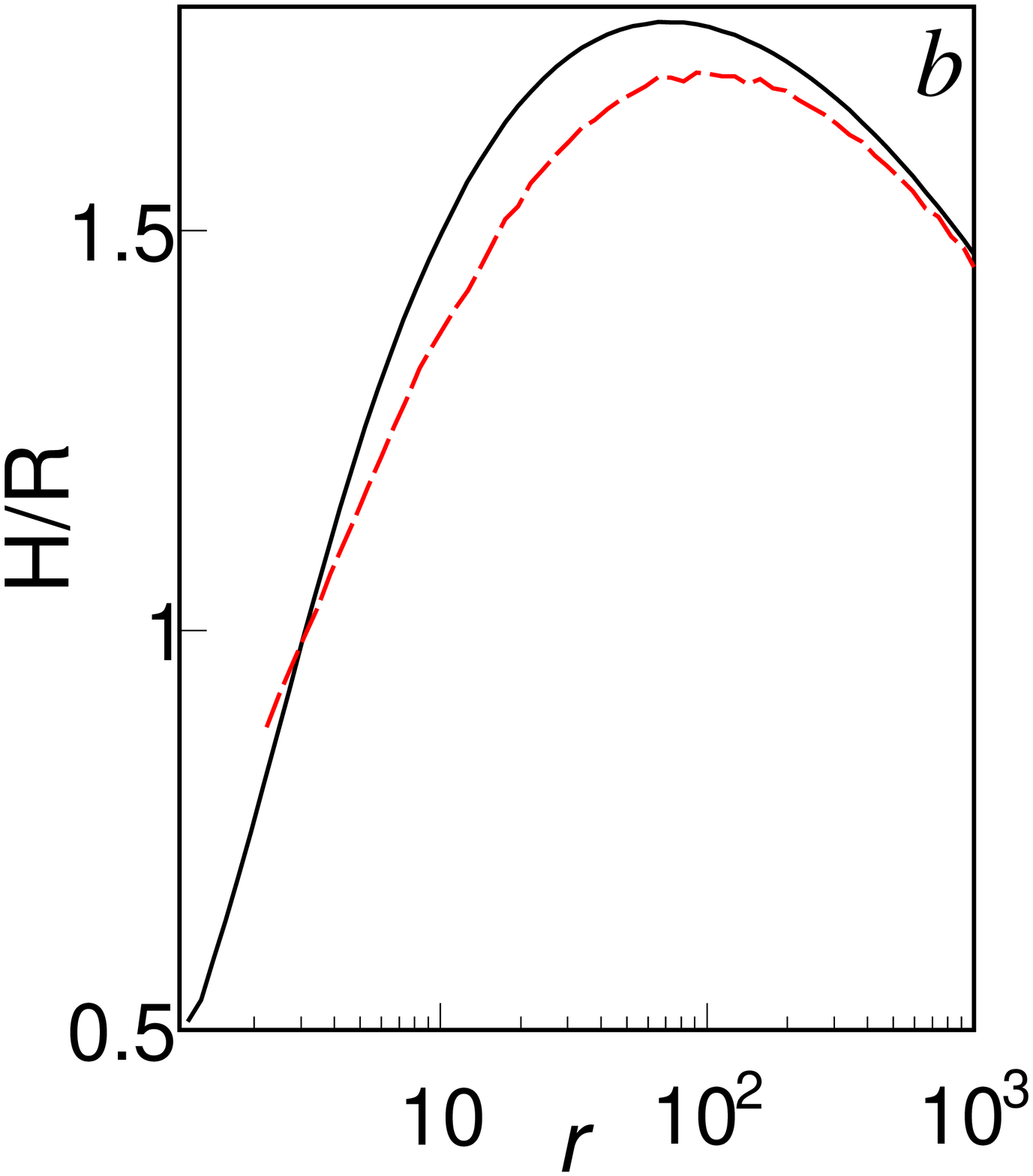}\hspace{0.12cm}
\includegraphics[width=4.2cm]{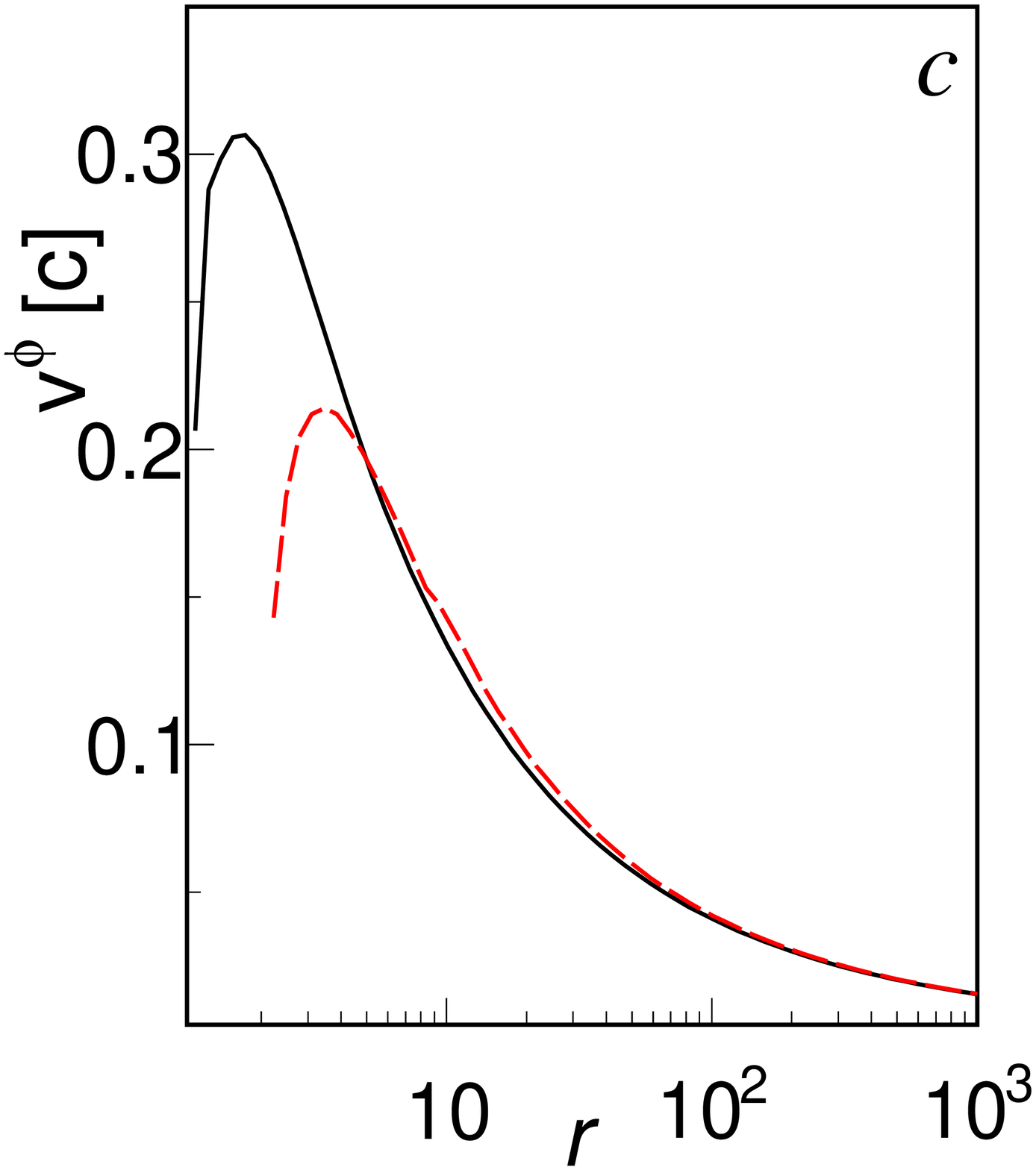}\hspace{0.12cm} 
 \includegraphics[width=4.2cm]{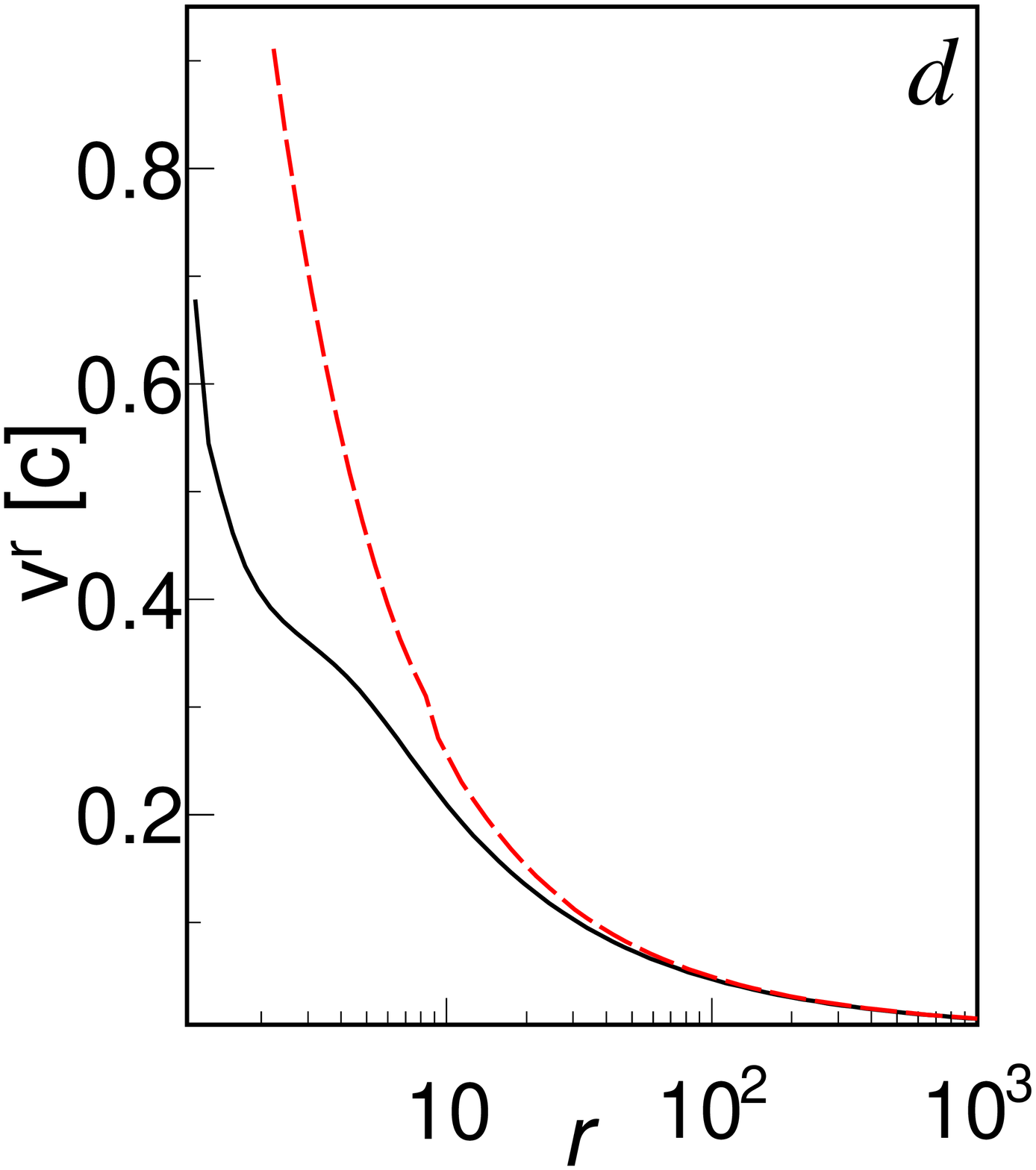}} 
 \caption{The radial profiles of hot flow parameters for $M =2 \times
 10^8\, \msun $ and $\dot m=0.1$. In all panels the solid (black
 online) and dashed (red) curves are for $a=0.998$ and $a=0$,
 respectively. (a) The vertical optical depth, $\tau_{z}
 \equiv n_e(r,0) \sigma_{\rm T} H (\pi/2)^{0.5}$. (b) The ratio
 of the scale height to the radius. (c) The azimuthal velocity. 
 (d) The radial velocity. These parameters are determined 
 in our initial (i.e., with the local slab approximation) solutions
 and are assumed to be the same in the final solutions. }
\label{fig:flow} 
\end{figure*}

\begin{figure*} 
\centerline{\includegraphics[width=4.2cm]{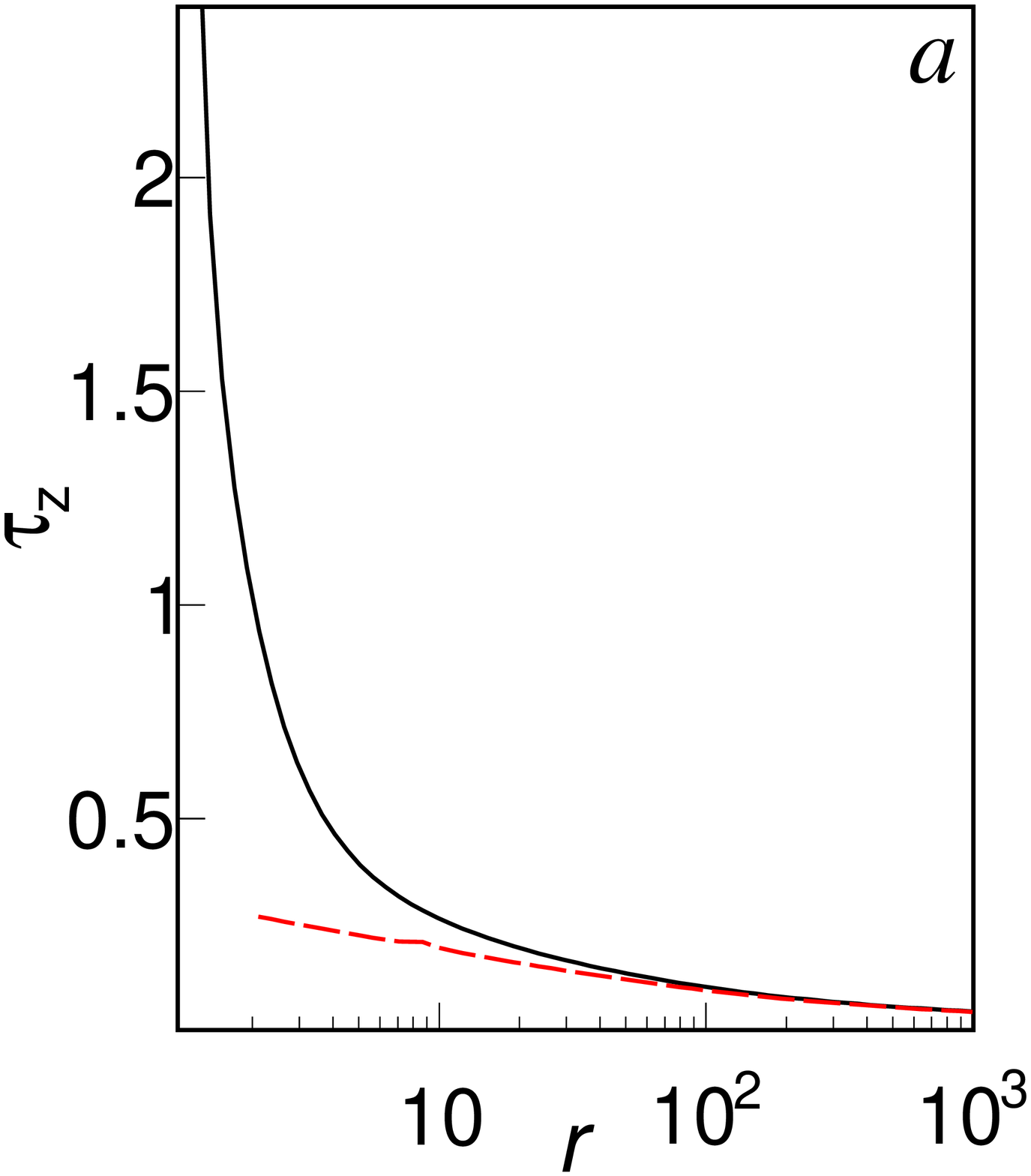}\hspace{0.12cm} 
 \includegraphics[width=4.2cm]{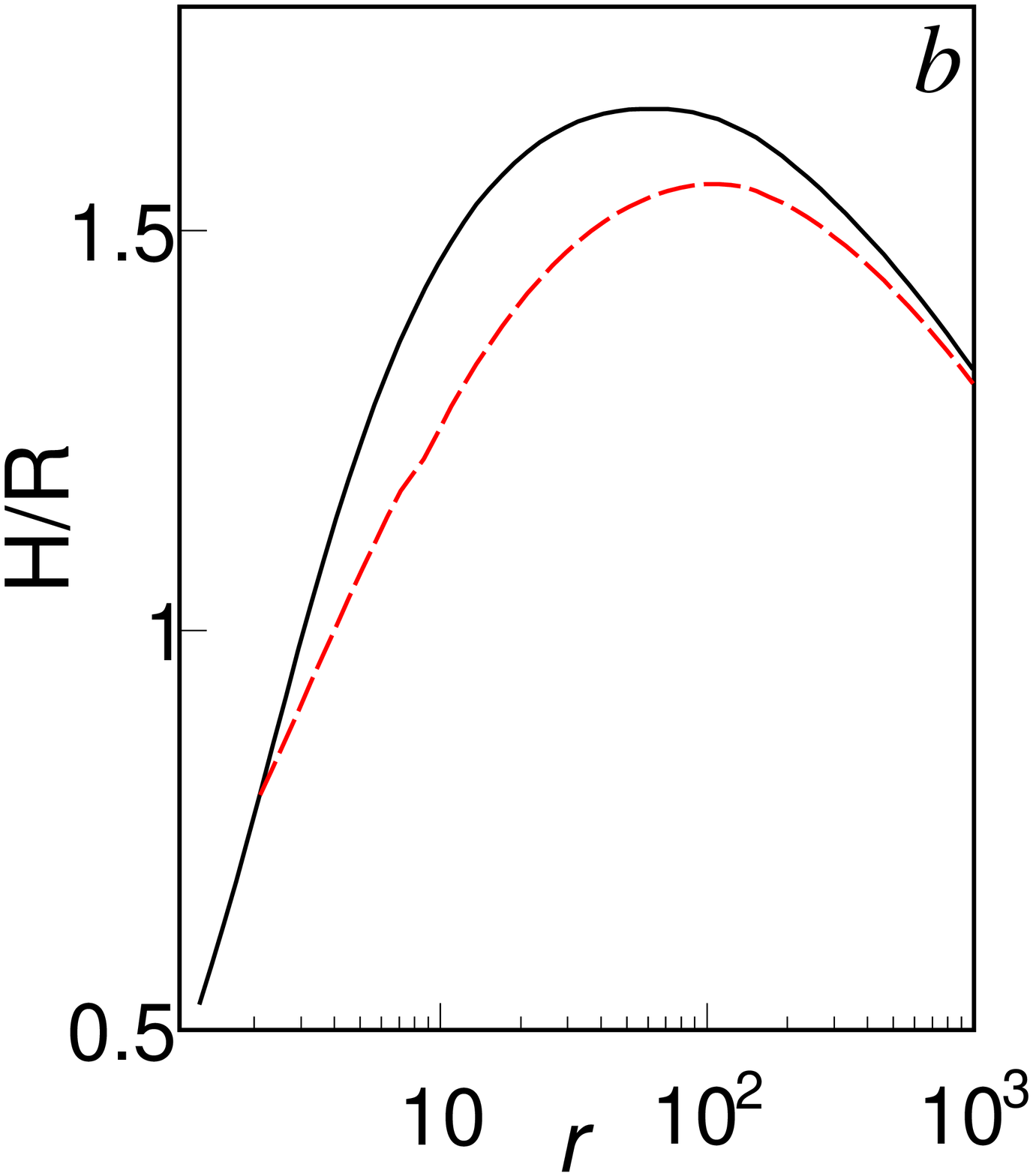}\hspace{0.12cm}
\includegraphics[width=4.2cm]{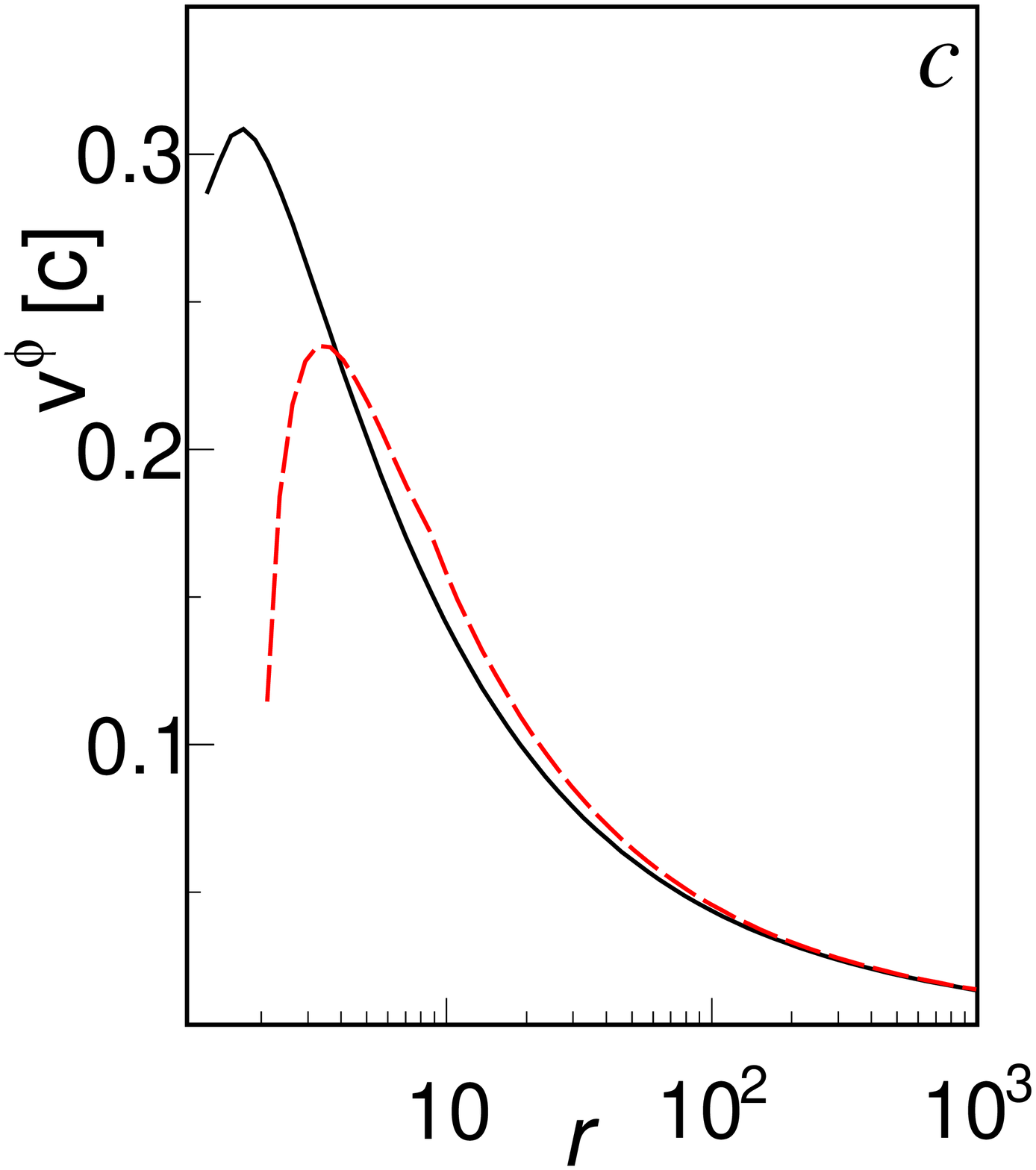}\hspace{0.12cm} 
 \includegraphics[width=4.2cm]{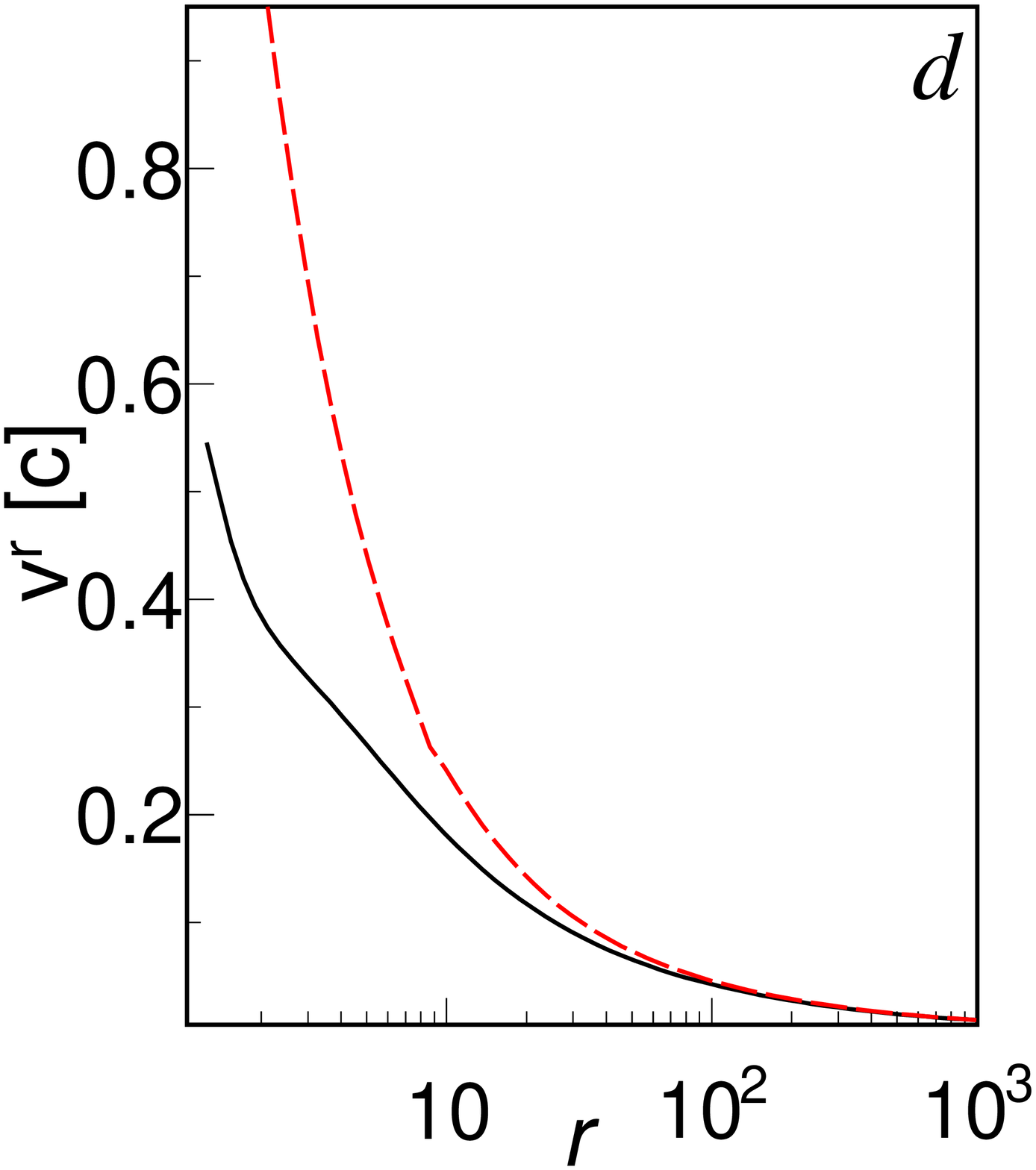}} 
\caption{Same as in Fig.\ 3 but for $M = 10\, \msun $ and $\dot
 m=0.5$. }   
\label{fig:flow_stel} 
\end{figure*}

Figs \ref{fig:rates} and \ref{fig:rates_stel} show the change of 
the heating and cooling rates, and the electron temperature, between 
our initial and final solutions. Figs \ref{fig:flow} and 
\ref{fig:flow_stel} show the remaining parameters of the flow
-- these are assumed to be the same in the initial and final
solutions. We can clearly see how the value of the spin parameter,
$a$, affects the flow properties. Rotation of the black hole
stabilizes the circular motion of the innermost part of the flow
(an effect directly related with properties of test particle motion,
analogous to the well-known dependence of the innermost stable orbit on $a$ in Keplerian discs) which yields a higher density (through
the continuity equation). This, in turn, implies both a stronger
Coulomb coupling (see Figs \ref{fig:rates}f and \ref{fig:rates_stel}f) 
and a larger optical depth (see Figs \ref{fig:flow}a and 
\ref{fig:flow_stel}a) in the innermost region in models with
$a=0.998$. On the other hand, heating by the compression work is
typically stronger than, or comparable to, heating by the Coulomb
energy transfer from ions (cf. Nakamura et al.\ 1997, Mahadevan \&
Quataert 1997), which effect partly reduces the differences between the high
and low $a$ models. These effects, and their impact on the
radiative properties of the flow, are discussed in detail below. 

Figs \ref{fig:rates}(f) and \ref{fig:rates_stel}(f) show also the dissipative
heating rate of ions, $Q_{\rm diss}$. The stabilized rotation results in a much stronger dissipative heating in the inner region for $a=0.998$, however,
this property is rather unimportant for effects investigated in
this paper (neglecting both the direct heating of electrons and
hadronic processes).

\subsection{Initial solutions} 
\label{initial} 
 
We discuss here in some details the initial solutions although they
are not self-consistent, because most of models available in literature
use the same local approximation for Comptonization. Moreover, we
amend some conclusions regarding the accuracy of local
approximations, derived previously in X10. We find the initial 
solutions using the prescription for the Compton cooling rate in a
slab geometry with initial photon energies of 1 eV, given in Dermer et
al.\ (1991; hereafter D91). In X10, we found that it gives a
reasonably good approximation for the innermost region. For the
models considered here, the slab approximation appears to be less
accurate, however, our choice of the slab case allows for a direct
comparison with previous studies, a number of which used the slab
approximation based on D91, or its modification introduced in Esin
et al.\ (1996). 

Figs \ref{fig:rates}(a,c) and \ref{fig:rates_stel}(a,c) show the
heating and cooling rates in the initial solution. Comptonization of
synchrotron photons is the most efficient radiative cooling process. 
However, the synchrotron emissivity decreases rapidly beyond its transition radius, $r_{\rm s}$ ($\simeq 20$ for $\dot m=0.1$ and $\simeq 100$ for $\dot
m=0.5$). Then, in the local model, the radiative cooling at
$r>r_{\rm s}$ is dominated by the (much weaker) bremsstrahlung and its
Comptonization. Then, two regions can be distinguished in terms 
of the {\it electron\/} energy equation (ions are always advection
dominated). (i) At $r>r_{\rm s}$, the flow is adiabatically 
compressed, which is reflected in the increase of $T_{\rm e}$
with decreasing $r$ (see Figs \ref{fig:rates}e and \ref{fig:rates_stel}e); both the radiative cooling and the Coulomb heating are much weaker than the advective
terms. (ii) At $r<r_{\rm s}$, the radiative cooling becomes
efficient and $T_{\rm e}$ decreases; the Coulomb heating is more
efficient in this region and it exceeds the compression work in the
innermost part (except for the model with $a=0$ and $\dot m =
0.1$). In this region, the advection of the internal energy contributes also
to the heating of electrons, however, this effect is rather weak. 

The solid curves in Figs \ref{fig:rates}(a,c) and 
\ref{fig:rates_stel}(a,c) show the GR global Compton cooling rate 
obtained in our initial MC simulations, i.e.\ with the $T_{\rm e}$ 
satisfying the energy equation (\ref{eq:energy}) with the Compton 
cooling rate, $Q_{\rm Compt,loc}$, given by the local slab (D91) 
prescription. As we can see, there are significant deviations between the local (dotted curves) and global cooling rates, resulting from several effects. Two of these effects should be generic to models of black-hole flows regardless of
specific values of parameters of the flow. First, the input of seed
photons from the inner region strongly enhances the Compton cooling
beyond $r_{\rm s}$. The second one involves the presence of the
event horizon (neglected in the local approximation). The presence of an inner 
boundary of the flow at the event horizon results in an obvious 
difference with respect to the semi-infinite slab; namely, a large fraction of photons generated in the innermost region is captured, reducing the input of seed photons. 

Another effect affecting the structure of the flow is the dependence of the optical depth from a given point on direction. This strongly depends on the flow parameters and assumptions about heating and outflow. In particular, the flows for our parameters have the optical depth from a point in the equatorial plane along the outward radial direction, $\tau_{r,{\rm out}}$, similar to the vertical one, $\tau_{z}$, at any $r$, whereas $\tau_{r,{\rm out}}\gg \tau_{z}$ in the model of X10. Furthermore, for $a=0.998$, both $\tau_{r,{\rm out}}$ and $\tau_{z}$ are much smaller than the optical depth in the inward radial direction, $\tau_{r,{\rm in}}$, except for the very innermost region; in these models $\tau_{r,{\rm in}} \simeq \tau_{r,{\rm out}}$ ($\simeq 0.2$ for $\dot m = 0.1$ and $\simeq 1$ for $\dot m = 0.5$) at $r \approx 2$. For $a=0$, $\tau_{r,{\rm in}} \simeq \tau_{r,{\rm out}}$ ($\simeq 0.05$ for $\dot m = 0.1$ and $\simeq 0.3$ for $\dot m = 0.5$) at $r \approx 10$; in these models the difference between $\tau_{r,{\rm in}}$ and $\tau_{r,{\rm out}}$ typically does not exceed a factor of 2. Then, for a non-rotating black hole, the local radiative properties would be more accurately described by the spherical geometry, while for a rapidly rotating black hole
the local properties are intermediate between the sphere and the slab. Then, the assumption of the slab geometry overestimates the initial cooling rate in all models considered here.  

Finally, as we note in X10, the D91 formula is accurate for only two values 
of seed photon energies, 1 eV and 1 keV, and we use here the lower one. 
With this value, the formula gives a reasonable approximation for flows around a stellar black hole, where seed photons have energies $\ga 1$ eV. For flows
around a supermassive black hole, however, with typical seed photon
energies of $10^{-2}$ eV (for $M =2 \times 10^8\, \msun$), it
underestimates the cooling rate. Specifically, for this $M$ we
found deviations by a factor of $\simeq 3$ between the formula of D91 for 1 eV and the results of our {\it local\/} slab MC simulation using the exact energy distribution of seed photons (given by the synchrotron and 
bremsstrahlung emission distributions; see X10 for the description of our local MC model).

\subsection{Self-consistent solutions}
\label{final}

The initial solutions are qualitatively similar at both values of
$\dot m$ considered here, the most significant dependence on $\dot
m$ involves the increase of the value of $r_{\rm s}$ with increasing
$\dot m$. In contrary, detailed properties of the self-consistent,
final solutions are more sensitive to $\dot m$. 

Figs \ref{fig:rates}(b,d) show the heating and cooling rates of 
the final solutions for $\dot m=0.1$. At $r>10$, a significant fraction of the power heating electrons is stored as their internal energy rather than radiated
away. This is qualitatively similar to the initial solution, although
$Q_{\rm rad}$ is larger at $r>10$ in the global cooling model and
hence $Q_{\rm int}$ is smaller than in the local cooling model;
therefore, the temperature gradient is smaller in the outer region in
our final solutions (see discussion below). We note that the strong
contribution of the $Q_{\rm int}$ term significantly complicates the
procedure for finding the self-consistent solution at this lower
$\dot m$, as both the global Compton cooling and the electron
advection are strongly sensitive to, and affect, the electron
temperature profile.  

Fig.\ \ref{fig:rates}(e) shows the electron temperature
profiles for the initial and the final solutions. We see that, in
the final solution, $T_{\rm e}$ increases compared to the initial
solution, by a factor of $\la 1.5$ in the inner region, and up to a
factor of 3 at several hundred $R_{\rm g}$. The increase of $T_{\rm e}$ 
at $r<20$ is an obvious effect, related with the
overestimation of the cooling rate by the slab approximation, combined
with - crucial in that region - capturing of seed photons by the
black hole, as discussed in Section \ref{initial}. 
Somewhat surprisingly, the global solution requires higher $T_{\rm e}$
also at $r>20$, where the global radiative cooling is significantly
enhanced. The temperature must remain high in this outer region
because the energy balance could not be achieved for $T_{\rm e}$
decreasing too rapidly. Namely, we found that such a (too rapid)
decrease of temperature at $r>20$ would yield a large positive $Q_{\rm
 int}$ term in the electron energy equation (\ref{eq:energy}), 
which, added to the (increased in global cooling) $Q_{\rm rad}$ term,
could not be balanced by the heating terms. We emphasise that this
property is related to strong contribution of advective terms and it
should not necessarily occur in the whole range of parameters, in
particular at higher accretion rates. 

For $\dot m=0.5$ (Fig.\ \ref{fig:rates_stel}), the self-consistent $T_{\rm e}$ is higher in the innermost part than $T_{\rm e}$ in the initial solution, due to the same reasons as for $\dot m=0.1$. However, the Compton cooling is relatively more important in the outer region, which results in an approximately 
isothermal structure at $r>4$; in particular, we do not find a local
maximum of $T_{\rm e}$, marking the transition (at $\simeq r_{\rm s}$) from an adiabatically compressed flow to an efficiently 
cooled flow in global solutions with $\dot m=0.1$ and in all our
local solutions. The temperature at which the energy balance is
achieved is twice lower for $a=0.998$, due to a much larger input
of seed photons from the central region.  

The strong contribution of the compressive heating has crucial consequences for the radiative efficiency, $\eta \equiv L/{\dot M c^2}$, where $L$ is the total observed luminosity of the flow, and for its scaling with $\dot m$. We compare here our models with the same $a$ for different $\dot m$, although we note that they have also different $M$, which may slightly affect such a comparison (the main dependence on the value of $M$ concerns the energy of synchrotron photons, influencing the cooling rate). Comparing our results for $\dot m=0.1$ and 0.5, we find that the fraction of an accretion power transferred to electrons via Coulomb collisions varies approximately as $\Lambda_{\rm ie,tot}/{\dot M c^2} \propto \dot m$ for $a=0.998$ and for $a=0$ we note even stronger dependence on $\dot m$. Then, heating dominated by the Coulomb coupling would lead to an approximately linear scaling of the radiative efficiency with $\dot m$, as derived in previous works on this subject (see, e.g., eq.\ 12 and references in Narayan \& McClintock 2008; see also table 3 in Rajesh \& Mukhopadhyay 2010). However, $\Lambda_{\rm compr,tot}/{\dot M c^2}$ is much less dependent, moreover negatively, on $\dot m$; specifically, it is higher by a factor of $\simeq 3$ for $\dot m=0.1$ than for $\dot m=0.5$. As a result, the dependence of $\eta$ on $\dot m$ is significantly reduced. 

For $a=0$, $\Lambda_{\rm compr,tot} \gg \Lambda_{\rm ie,tot}$ 
at $\dot m=0.1$ and $\Lambda_{\rm compr,tot} \ga \Lambda_{\rm ie,tot}$ 
at $\dot m=0.5$. The decrease of $\Lambda_{\rm compr,tot}/{\dot M c^2}$ with increasing $\dot m$ is partially balanced by the increased contribution of Coulomb heating. Two further effects decrease the radiative efficiency at lower $\dot m$: (1) the increase of the fraction of the energy given to electrons which is not radiated away (and increases the internal energy advected with the flow), and (2) the increase of the fraction of photons captured by the black hole (Section \ref{collimation}). As a result, the radiative efficiency for $a=0$ is approximately the same for both considered values of $\dot m$. More specifically, $\eta \approx 0.0042$ (with $L=1.1 \times 10^{43}$ erg s$^{-1}$ and $L/\ledd =4.2 \times 10^{-4}$) for $\dot m=0.1$ and $\eta \approx 0.0045$ ($L=2.9 \times 10^{36}$ erg s$^{-1}$ and $L/\ledd=2.3 \times 10^{-3}$) for $\dot m=0.5$. 

For $a=0.998$, $\Lambda_{\rm ie,tot}$ exceeds $\Lambda_{\rm compr,tot}$ at both $\dot m=0.1$ and 0.5. Radiative efficiencies at these $\dot m$ differ only by a factor of 2, due to relatively stronger contribution of compressive heating at $\dot m=0.1$ and very strong reduction of the observed $L$ by GR effects (Section \ref{collimation}) at $\dot m=0.5$. Specifically, for $\dot m=0.1$, $\eta \approx 0.0078$ ($L = 2 \times 10^{43}$ erg s$^{-1}$ and $L/\ledd =7.8 \times 10^{-4}$) and for $\dot m=0.5$, $\eta \approx 0.016$ ($L = 1.1 \times 10^{37}$ erg s$^{-1}$ and $L/\ledd=8.3 \times 10^{-3}$). We note that for $a=0.998$ the apparent luminosity depends on $\theta_{\rm obs}$, see Section \ref{anisotropy}, and the above values of $L$ are averaged over all viewing angles. 

\begin{figure} 
\centerline{\includegraphics[width=8cm]{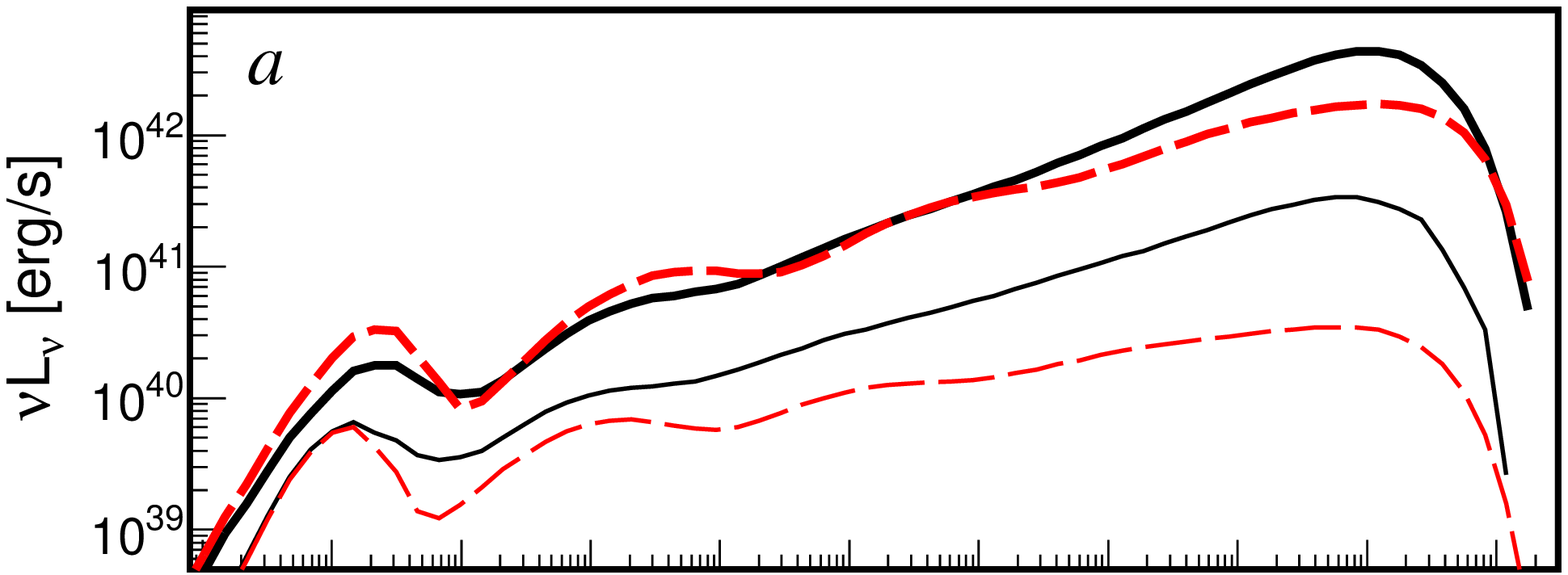}} 
\centerline{\includegraphics[width=8cm]{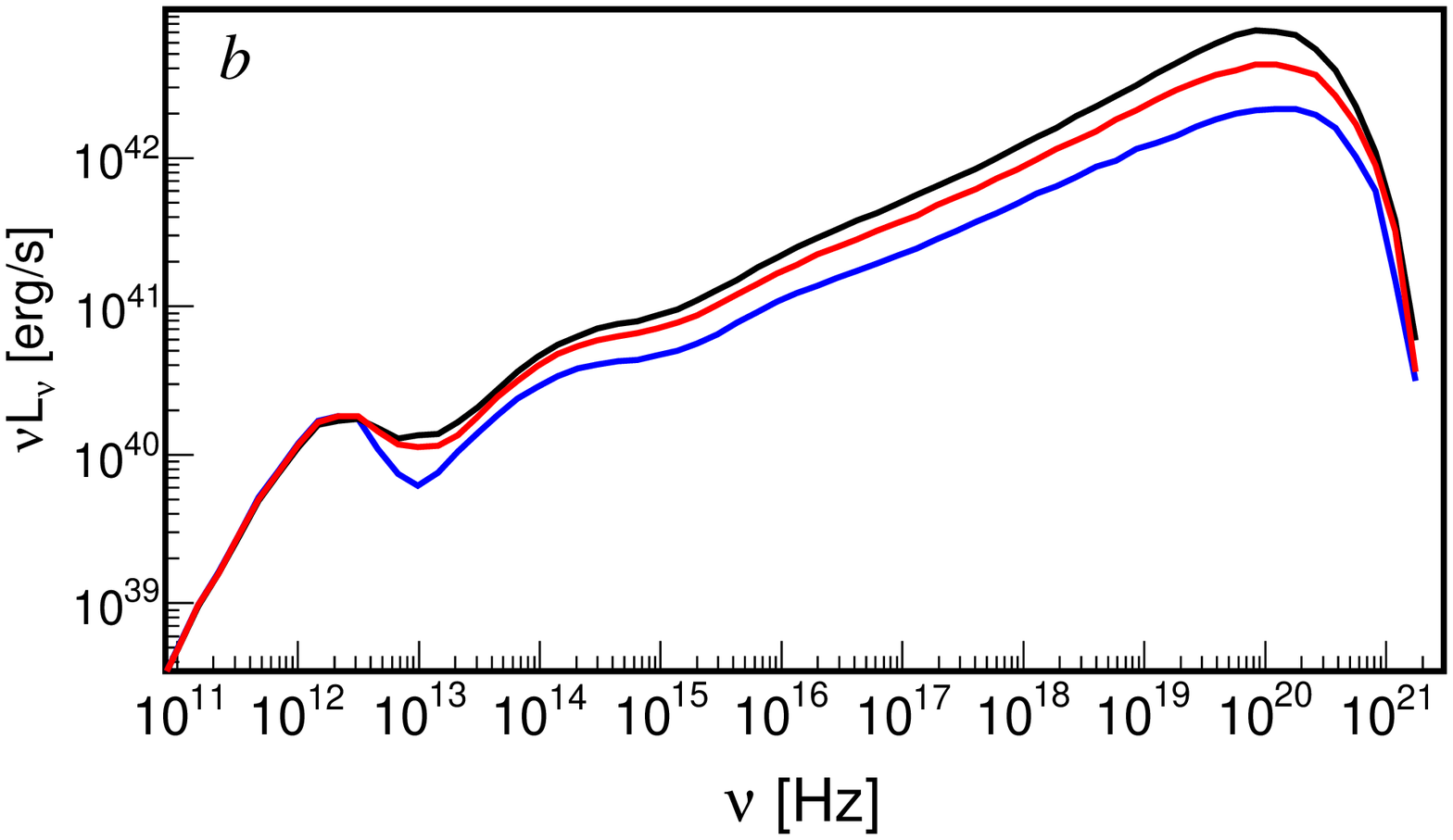}} 
\caption{(a) The angle-averaged spectra from the MC simulations for
 the initial (thinner curves) and final (thicker curves) solutions in
 models with $M =2 \times 10^8\, \msun $ and $\dot m=0.1$. The
 solid (black) and dashed (red) curves are for $a=0.998$ and $a=0$, 
 respectively. (b) The angular dependence of spectra for the final
 solution with $a=0.998$; the spectra from bottom to top are for
 $\mu_{\rm obs}=0.9$--1, 0.5--0.6 and 0--0.1. } 
\label{fig:spectra} 
\end{figure}

\begin{figure} 
\centerline{\includegraphics[width=8cm]{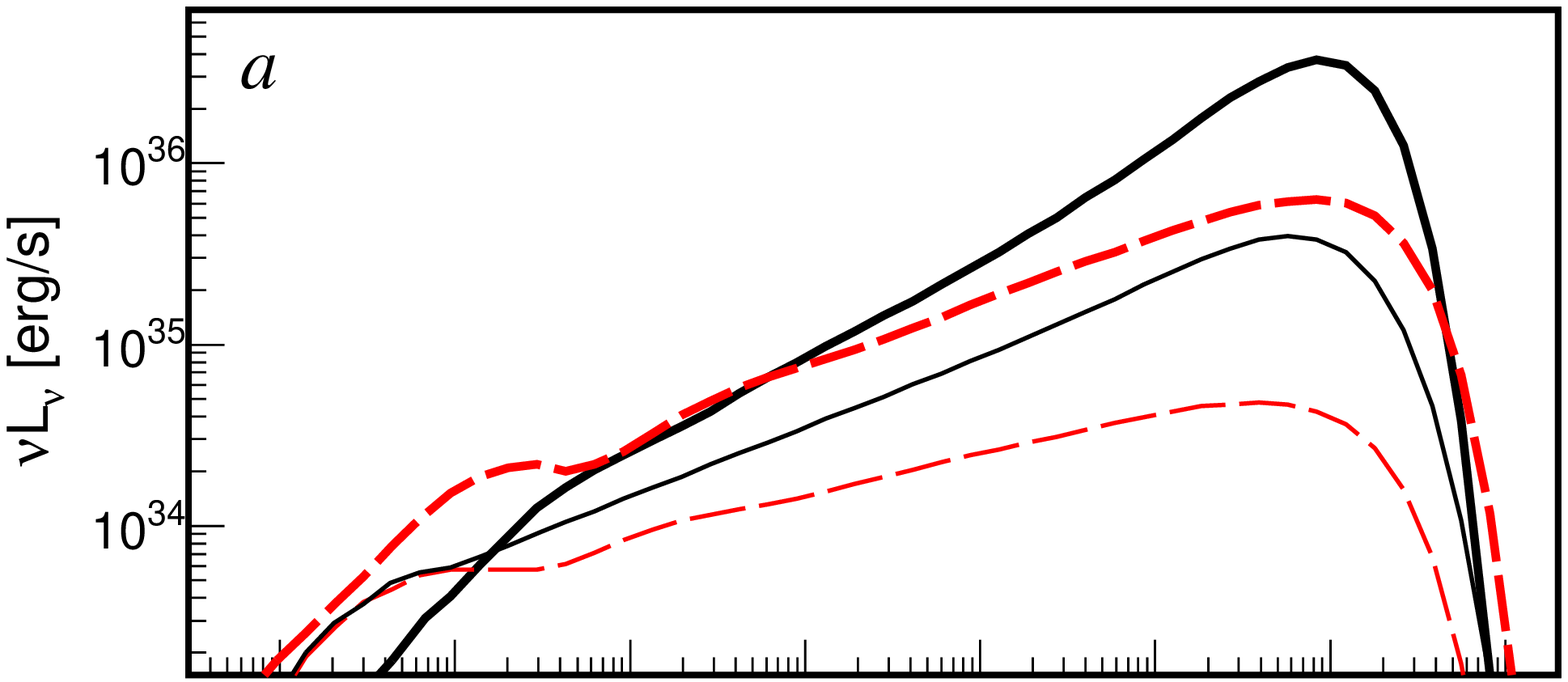}} 
\centerline{\includegraphics[width=8cm]{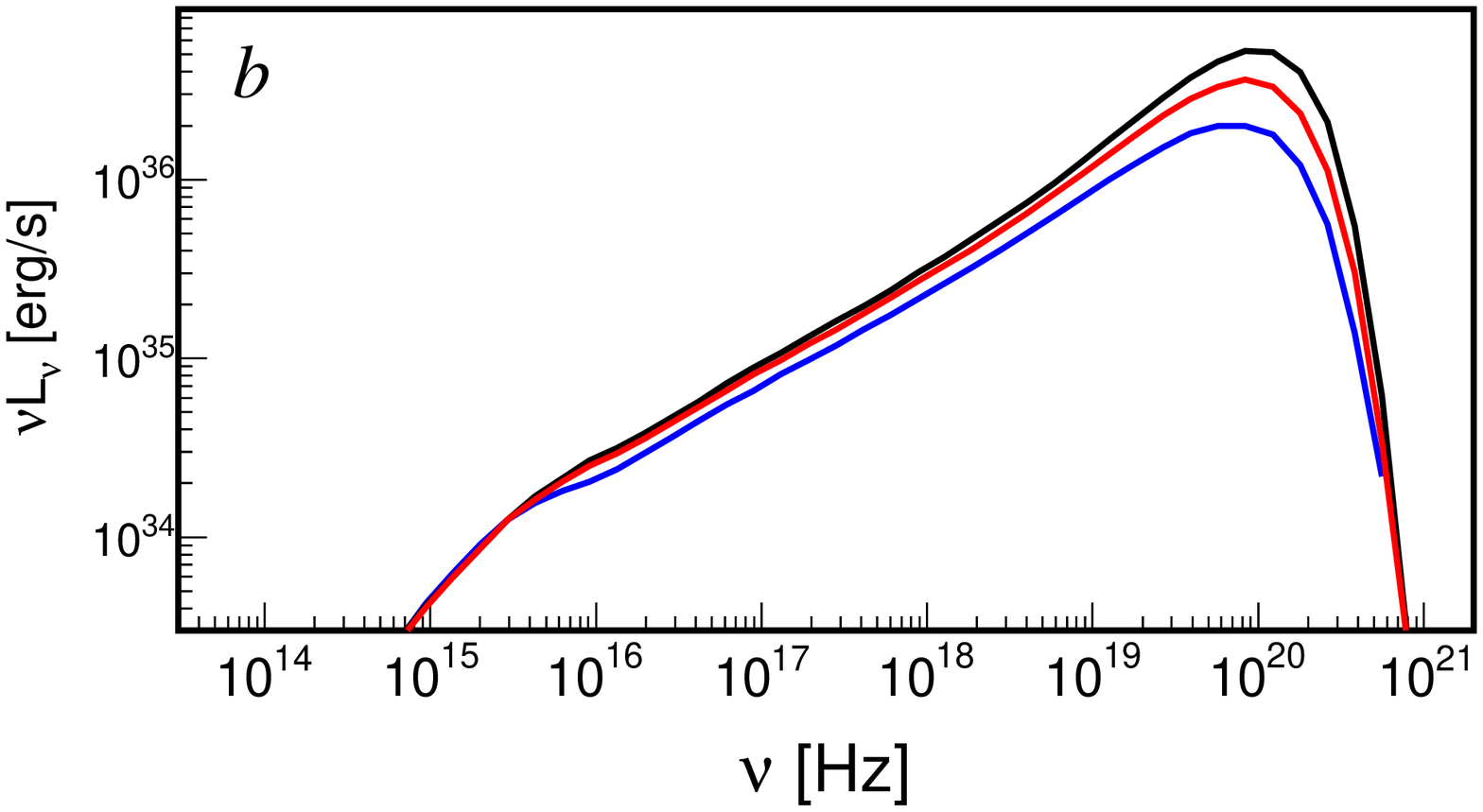}} 
\caption{Same as in Fig.\ \ref{fig:spectra} but for $M = 10\, \msun$ 
and $\dot m=0.5$. } 
\label{fig:spectra_stel} 
\end{figure}

Figs \ref{fig:spectra}(a) and \ref{fig:spectra_stel}(a) show the 
angle-averaged spectra for the initial and final solution. In all models, the normalization of the spectra of the final solution is much higher than that of the initial solution, due to the increase of the synchrotron emissivity 
resulting from the increase of $T_{\rm e}$. The X-ray
spectra are relatively hard, with the photon index $\Gamma \approx 1.6$,
for both ($\dot m=0.1$, $a=0.998$) and ($\dot m=0.5$, $a=0$) and $\Gamma \approx 1.5$ for ($\dot m=0.5$, $a=0.998$); the last value is determined in the 2--10
keV range, as at higher energies a rather pronounced Wien peak leads
to further hardening of the spectrum. For ($\dot m =0.1$, $a=0$),
the relativistic $T_{\rm e}$ and small optical depth  result in
pronounced scattering bumps seen in this model spectrum; the average slope in the 2--200 keV range is $\Gamma \approx 1.7$. Thus, we see that the spectra harden with the increasing luminosity, for $\dot m$ increasing from 0.1 to 0.5, from $\Gamma\simeq 1.7$ to $\simeq 1.6$ at $a=0$ and from $\Gamma\simeq 1.6$ to $\simeq 1.5$ at $a=0.998$.

In all models, the $EL_E$ spectra have the maxima at several hundred keV. The electron temperature reaches the maximum values of $\sim 1$ MeV for $\dot m=0.1$ and $\sim 400$ keV for $\dot m=0.5$. 

Note, however, that although these flows produce photons with energies $>m_{\rm e} c^2$, we do not expect an efficient e$^\pm$ pair production in
photon-photon collisions. The probability that a $\gamma$-ray produces a pair is approximately given by the compactness parameter for this process, $\ell(R) \equiv L_{X,R} \sigma_{\gamma \gamma} / (4 \upi m_{\rm e} c^3 R)$, where $L_{X,R}$ is the radiative power in photons with $E > 10$ keV produced within $R$. It is $\ll 1$ for all $R$ in all models except for $\dot m=0.5$ and
$a=0.998$, where $\ell \ga 1$ in the innermost few $R_{\rm g}$. The
same conclusion was reached in previous studies of pair equilibria in
advection-dominated flows (e.g.\ Esin 1999) where additional pair
creation processes, namely electron-electron and photon-particle
collisions, were also taken into account.  

\subsection{Effects of black hole rotation}

The nature of the space-time metric (specified by the
value of $a$) affects the radiative properties of hot flows through 
various effects. As noted above, the stabilized circular motion 
results in a much larger density in the innermost region in models
with large $a$. This affects the radiative efficiency in a manner
discussed in Section \ref{final}; here we restate some of these results with
the emphasis on the dependence on $a$. Two further effects,
discussed in Sections \ref{collimation}--\ref{anisotropy}, concern the properties of photon motion in curved space-time.  

\subsubsection{Heating}
 
Higher density implies a stronger Coulomb coupling for large $a$. 
For $a=0$, $\Lambda_{\rm ie,tot}$ is smaller than for $a=0.998$ by a
factor of 200 and 35 for $\dot m=0.1$ and 0.5, respectively. The
difference between the heating of electrons for high and low $a$ is,
however, strongly reduced by the high efficiency of compression,
which shows very weak dependence on the value of $a$. Then, the
total heating, $\Lambda_{\rm ie,tot} + \Lambda_{\rm compr,tot}$, is
higher in models with $a=0.998$ only by a factor of 3.5 ($\dot
m=0.1$) and 16 ($\dot m=0.5$).  Extrapolating our results we can
expect that at $\dot m \ll 0.1$ the difference between the heating
efficiencies in high and low $a$ models vanishes. 

On the other hand, a much larger difference between the heating rates for large and small values of $a$ could be expected in models involving the
direct transfer of the dissipated energy to electrons; compare
$\Lambda_{\rm ie}$ and $\Lambda_{\rm compr}$ with $Q_{\rm diss}$
shown in Figs \ref{fig:rates}(f) and \ref{fig:rates_stel}(f).  

\subsubsection{Optical depth}

Higher density implies also a larger optical depth for large $a$.
Then, despite a stronger heating in these models, the energy balance
in the innermost region is achieved at smaller electron temperatures
than for small $a$. On the hand, the highly efficient Comptonization
occurs at very small $r$, therefore, a much more centrally
concentrated emissivity of Comptonized radiation is achieved in
models with $a=0.998$ and the {\it observed\/} luminosity is subject to
stronger reduction by GR effects, discussed below. 
 
\subsubsection{Collimation and capturing of radiation}
\label{collimation}

Different velocity fields result in a different strength of
collimation toward the black hole horizon. In principle, larger
values of the radial velocity, $|v^r|$, imply a stronger collimation for $a=0$, which could give rise to a larger fraction of photons being captured by
the black hole. However, the radiation is produced, on average, at
a smaller radial distance for $a=0.998$ and, therefore, the
reduction of the observed luminosity due to GR effects, including the
photon capture and the gravitational redshift, turns out to be 
larger in that case. Specifically, in models with $a=0.998$,
$L/Q_{\rm rad,tot} \approx 0.2$ for $\dot m=0.5$ and 0.3 for $\dot
m=0.1$; the large reduction for $\dot m=0.5$ results from $\tau > 1$
(and trapping of photons) in the central region. For $a=0$,
$L/Q_{\rm rad,tot} \approx 0.7$ (for $\dot m=0.1$) and 0.9 (for $\dot
m=0.5$; in this case the emissivity profile is flatter). 

\subsubsection{Intrinsic anisotropy}
\label{anisotropy}

For high values of $a$, effects unique for the Kerr metric, i.e.\ bending of
photon trajectories to the equatorial plane combined with a
dependence of the gravitational shift of energy on the direction of
photon escape, result in an intrinsic anisotropy of radiation
produced within the innermost few $R_{\rm g}$ (see Piran \& Shaham
1977, Nied\'zwiecki 2005). The Comptonized component from that
region is softer (i.e.\ has a larger spectral index and a smaller
cut-off energy) and has a smaller normalization when observed at
a smaller $\theta_{\rm obs}$. The gravitational effects, underlying
this property, are strong only within $4R_{\rm g}$, then, the
magnitude of the anisotropy in the total spectrum depends on the
proportion between contributions from the innermost and the
surrounding (beyond $4 R_{\rm g}$) regions. 

Figs \ref{fig:spectra}(b) and \ref{fig:spectra_stel}(b) show the $\theta_{\rm obs}$-dependence of the observed spectra in our models with $a=0.998$ 
(in models with $a=0$ the dependence on $\theta_{\rm obs}$ is negligible). 
As we can see, the spectra indeed show some anisotropy but the effect is relatively moderate. For $\dot m=0.1$, the total flux emitted face-on ($\mu_{\rm obs}=0.9$--1) and edge-on ($\mu_{\rm obs}=0$--0.1) corresponds to the isotropic
luminosity of $1 \times$ and $3 \times 10^{43}$ erg s$^{-1}$,
respectively. In this model, the compressive heating results in a
rather flat emissivity profile, then, mixing of relatively weak
contribution from $r<4$ with radiation produced at more distant
regions washes out the anisotropic properties. For $\dot m=0.5$,
the face-on and edge-on fluxes correspond to $6.6 \times 10^{36}$
erg s$^{-1}$ and $1.5 \times 10^{37}$ erg s$^{-1}$, respectively. In this model,
the emissivity is strongly centrally concentrated; however, the
optical depth of the inner region is large and a large fraction of
photons produced there get reprocessed at larger distances or
captured. At both $\dot m$, the spectral indices differ by
$\Delta \Gamma \la 0.1$ between a face-on and an edge-on observer. 

We expect that a much larger anisotropy would be
produced in a model with moderate $\dot m$ (yielding $\tau<1$) 
and strong direct heating of electrons (giving a very centrally
concentrated emissivity). 
   
\subsection{Illumination by an outer disc}
\label{outer}

An optically thick disc may extend to relatively low radii in some objects observed at $L/\ledd$ similar to these characterising our
models. The best example concerns XTE J1118+480, in which the derived truncation radius of the outer cold disc is $\simeq 100 R_{\rm g}$ (Esin et al.\ 2001).
We consider here the cooling of the inner hot flow due to Comptonization 
of thermal photons emitted by such a surrounding, cold disc.
For this we take the innermost part, within $r = 100$, of our self-consistent solutions described in Section \ref{final}. We use our MC method to compute the Comptonization 
of the seed photons emitted by a cold, Keplerian disc extending 
from $r_{\rm in} = 100$ to $r_{\rm out} = 1000$. The local black-body 
temperature of its thermal emission is found using Page \& Thorne (1974); note that this approach gives the largest possible luminosity of the outer
disc as it involves the assumption that the rate of the outward transport
of mechanical energy, which is the dominant source of the radiated energy  
at $r > 100$, is the same as in an untruncated Keplerian disc.
We do not look for self-consistent solutions including this additional source of seed photons. Instead, we compute the total rate of the Compton cooling by the outer disc photons, $Q_{\rm Compt,tot}^{\rm th}$, and compare it with the total Compton cooling rate due to the synchrotron and bremsstrahlung emission,  $Q_{\rm Compt,tot}$, of our self-consistent solutions.
 
We find that in models with $a=0$ the effect is small, with 
$Q_{\rm Compt,tot}^{\rm th}/Q_{\rm Compt,tot} \approx 0.2$,
and for $a=0.998$ it is negligible, 
$Q_{\rm Compt,tot}^{\rm th}/Q_{\rm Compt,tot} < 10^{-2}$.
We note that the weakness of this effects results from the following 
properties. First, the total luminosity of the disc truncated at $r_{\rm in}=100$ is relatively small (specifically, it is approximately equal to the luminosity of the hot flow). Second, the solid angle subtended by the flow as seen by the outer disc is rather small. 
Third, the optical thickness of the outer parts of the flow is very small
at the values of $\dot m$ considered here, so a small fraction of photons illuminating the flow get scattered.

Our results are consistent with the conclusion of Esin et al.\ (2001) that the outer disc is not an important source for Compton cooling in this range of parameters. Obviously, the effect would be more important for a smaller truncation radius (cf.\ Esin 1997), at which the luminosity of the disc would be larger, as well as at higher $\dot m$, at which a larger fraction of photons 
irradiating the flow would be scattered.

\subsection{Comparison with previous works}
\label{previous}

Most of the specific properties discussed in this paper
have been studied in previous works. However, their quantitative
importance could not be assessed properly as neither of previous
studies considered self-consistently all relevant effects. 

The solution of a hot-flow structure based on a self-similar 
model (e.g., Narayan \& Yi 1995) is commonly used due to its relative simplicity. However, this simple solution introduces inaccuracies of up to an order of magnitude even for the basic flow parameters (e.g., the density given by eq.\ 1 in Mahadevan \& Quataert (1997) is $\sim 10$ times larger than the
density in our solutions), which leads to further inaccuracies in
papers based on this model; e.g., we find that compressive heating
dominates over Coulomb heating at much higher accretion rates than
the critical value given by eq.\ (35) in Mahadevan \& Quataert (1997). 

Gammie \& Popham (1998) and Popham \& Gammie (1998)
presented an extensive study of various aspects of GR
hydrodynamical description of a flow in the Kerr metric and they show
that the black hole spin has a large effect on the velocity field,
density and temperature of the flow (in ways also discussed in the present
paper); however, their study neglected radiative processes, so a
more detailed impact on the observed spectrum could not be established. 

Manmoto, Mineshige \& Kusunose (1997) included an explicit description of radiative processes with local approximation of Compton cooling in a pseudo-relativistic model. The same model including outflows and direct electron heating was used by Yuan et al.\ (2003, 2005). Then, M00 improved such models by including a fully GR treatment of the hydrodynamical processes (but still with local approximation of Compton cooling). We have fully
reproduced their results in our initial solutions; the accuracy of
these models is directly illustrated by comparison of our initial
and final (global) solutions.  

In an alternative (to MC) method for the treatment of global Comptonization
in an optically thin flow, developed by Narayan, Barret \& McClintock (1997),
the flow is divided into a set of nested spherical shells and the iterative scattering method is used to describe their mutual interactions. The model was used, e.g.\ by Esin, McClintock \& Narayan (1997), Quataert \& Narayan (1999) and Esin et al.\ (2001). Neither of these papers presents any details which could allow us to compare our cooling rates and only the produced spectra can be compared. The results of Esin et al.\ (2001), who use the GR hydrodynamical model from Gammie \& Popham (1998) and include the special relativistic effects and the gravitational redshift in the calculation of radiative processes, is of particular interest for such a comparison. It appears that the model of Narayan et al.\ (1997) yields a higher cooling rate than our Monte Carlo model, as the spectrum computed in Esin et al.\ (2001) for parameters similar to these in our models (see the solid curve in their fig.\ 1) has a break energy at $\sim 100$ keV, significantly lower than our model spectra. We do not know the cause of this difference.

Among previous studies of hot flows, the fully GR ADAF model of Kurpiewski \& Jaroszy\'nski (1999, 2000), with a MC computation of global Comptonization, seems to be the closest to our present work. Their spectra show a much
stronger dependence on $a$, see e.g.\ figure 4 in  Kurpiewski \& Jaroszy\'nski (2000), which may be due to their values of $\dot m$ significantly smaller than in our models. 

Yuan et al.\ (2009) used an iteration method, similar to ours but with a different approach to computing global Comptonization, to find the radiative cooling rate mutually consistent between the dynamics and radiation of the flow in a non GR model including strong outflows and electron heating.
Taking into account significant differences between the results of our present work and X10, pointed out below, we speculate that GR flows with a different strength of these two effects may be characterised by different critical accretion rates, and different maximum luminosities, than those found in the model of Yuan et al.\ (2009). 

Finally, we compare our results with the pseudo-relativistic model of X10. First, we note that models based on the pseudo-Newtonian potential fail to
properly describe the innermost region even for the $a=0$ case. 
Specifically, the pseudo-relativistic model predicts a decrease of the
optical depth at $r<7$ (see the dot-dashed curve in fig.\ 2b in
X10) while the fully GR model with the same parameters
(see our Fig.\ \ref{fig:flow}a) yields the optical depth monotonically increasing with decreasing $r$. The reason for this discrepancy can be traced to the (unphysically) large radial velocity in the pseudo-relativistic model, formally exceeding $c$ at small $r$, which implies a much smaller density than that in the GR model.  

Second, in the present work we use the equation for vertical equilibrium 
in the form derived for the Kerr metric by Abramowicz, Lanza \& Percival (1997).
In X10 we used its simplified version, $H = c_{\rm s}/\Omega_{\rm K}$,
where $c_{\rm s}$ is the sound speed and $\Omega_{\rm K}$ is the Keplerian angular velocity,
and we note significant differences in scale heights derived in these
two approaches. Specifically, the equation from Abramowicz et al.\ (1997)
yields a larger scale height, up to a factor of $\sim 2$, for the flow parameters
describing our models.

Third, additional effects included in X10, namely the strong outflow 
and direct viscous heating of electrons affect the flow structure in a complex manner and lead to a qualitatively different effect of taking into account the global nature of the Comptonization process. Obviously, the outflow leads to a reduction of $\tau_z$ at small $r$. Therefore, $\tau_z$ is much smaller,
by a factor of $\sim 10$ at $r<10$, in the model of X10 than in our present
model with $a=0$ and $\dot m=0.5$ (which parameters are the same as those in X10). Then, the outflow reduces the compressive heating of ions via reducing the density gradient, and the direct heating of the electrons reduces the viscous heating of ions. Also, the outflow reduces the viscous heating rate per unit volume, flattening its radial profile, but the viscous heating rate per ion remains unchanged. We note that the compressive heating of ions exceeds their viscous heating in our models, and it is much larger, by a factor of $T_{\rm i}/T_{\rm e}$, than the compressive heating of electrons. Together, both reductions of the ion heating cause a reduction of $T_{\rm i}$, which, in turn, leads to a decrease of $H/R$ by $\sim 30$ per cent (with respect to the pseudo-relativistic model without both an outflow and electron heating) within the innermost several
$R_{\rm g}$. 

The simplified hydrostatic equilibrium condition together with the outflow and electron heating considered in X10 yield a much smaller $H/R$ than our present model (compare our Fig.\ \ref{fig:flow_stel}b with fig.\ 3b in X10). As a result, in spite of the strong mass loss to the outflow, the total {\it radial} optical depth in the model of X10 is still higher than that in our model with $a=0$ and $\dot m=0.5$. Also, the radial optical depth of X10 is much larger than $\tau_z$; then, the geometry of the flow resembles that of a slab. 

Fourth, comparing the radiative efficiencies of the X10 model and the present one with $a=0$, $\dot m=0.5$ illustrates the strong impact of the direct heating of electrons. Namely, in X10 the value of $\eta$ is three times larger, although most of the potential energy of the flow is lost to the outflow in their model.

\subsection{Strength of magnetic field and electron heating}
\label{delta}

We discuss here briefly our assumption on the strength of the magnetic field as well as our neglect 
of the viscous heating of electrons. Our assumption of rather weak
magnetic field, with the magnetic pressure of $0.1p$, is supported by results of the magnetohydrodynamic (MHD)
simulations in which amplification of magnetic fields by the MRI typically saturates at such a ratio of the magnetic to the total pressure (e.g.\ Machida, Nakamura \& Matsumoto 2004 and references therein). It is rather easy to assess changes of the flow structure and the spectrum resulting from a change of this assumption. Namely, a stronger magnetic
field would result in a stronger synchrotron emission and this would cause a decrease of the electron temperature
and a softening of the Comptonization spectrum (see, e.g., Esin et al.\ 1997).

An issue of the direct viscous heating of electrons is more uncertain.
Current applications of ADAF models typically assume large values of electron heating to total heating,
$\delta \sim  0.5$ (e.g., Yuan et al.\ 2003), however, the model has
a degeneracy between the value of $\delta$ and the strength of an outflow (Quataert \& Narayan 1999).
Various attempts to assess this effect on theoretical grounds seem to 
favour small values of $\delta$ for weak magnetic fields  
and rather high $\dot m$, which are considered in our models.
Analytic investigation of particle heating by MHD turbulence in ADAFs
indicates that the turbulence primarily heats protons for weak magnetic fields,
while electrons are primarily heated for strong magnetic fields (see Quataert \& Gruzinov 1999).
However, large uncertainties in the division between electron and proton heating result from
uncertainties in the description of the turbulence; $\beta_{\rm B}=0.9$, assumed in our models, may correspond
to the values of $\delta$ between $\simeq 0.006$ and 0.6 (see fig.\ 2 in Quataert \& Gruzinov 1999). 
Details of another potential mechanism of electron heating discussed by Bisnovatyi-Kogan \& Lovelace (1997)
and Quataert \& Gruzinov (1999), i.e.\ reconnection,
remain even more uncertain. 

Simulations of MRI turbulence by Sharma et al.\ (2007)
indicate that pressure anisotropy, created in turbulent plasmas,
may give an additional mechanism for particle heating; they approximate the fraction 
of the viscous energy which heats electrons as $\delta \approx (T_{\rm e} /T_{\rm i})^{1/2}/3$.
However, their model assumes a fully collisionless plasma, which approximation is valid only for
$\dot m \ll 0.06$. At higher accretion rates, Coulomb collisions suppress the electron pressure anisotropy
and electron heating is negligible. Then, their prescription for $\delta$ seems to be not relevant for our models.

We note that the direct viscous heating would exceed the Coulomb and compressive heating
of electrons for $\delta \ga 0.1$ in models with  $a=0$, and for $\delta \ga 0.01$ in models with  $a=0.998$.
For these values of $\delta$, flows should have larger luminosities 
than these assessed in our models. We can also expect that flows with such values of $\delta$ produce 
harder X-ray spectra, with higher cut-off energies, than obtained in our models.
However, we note also that large values of $\delta$, for which the heating of ions 
is significantly reduced, may have a strong impact on the flow structure, especially 
if a strong outflow is also present, which may lead to enhanced global Compton
cooling (as discussed in Section \ref{previous}). Then, the above simple estimation
of spectral changes related to the increase of $\delta$ may be incorrect; in particular, we note
that the spectrum obtained in our model with $\delta=0.5$ in X10 has a smaller
cut-off energy than our spectra computed here with $\delta=0$.

\section{Comparison with observations} 
\label{comparison}

\subsection{FR I radio galaxies}
\label{fr1}

A substantial observational evidence indicates that accretion
flows typically proceed through a radiatively-inefficient mode
below a characteristic luminosity of $\simeq 0.01\ledd$ in AGNs
(e.g., Ho 2008; see Yuan 2007 for a review on the applications of
hot flow models to low luminosity AGNs). FR I radio galaxies make an
interesting class of objects observed at such luminosities. They do
not show signs of the presence of an optically thick accretion disc
(e.g., Chiaberge, Capetti \& Celotti 1999). Remarkably, an average
radiative efficiency, $\eta \la 0.005$, estimated for FR Is by
Balmaverde, Baldi \& Capetti (2008) is consistent with the values of
$\eta$ obtained in most of our solutions. 

According to our results, hot flows with relevant accretion rates
should produce rather hard spectra, extending to MeV energies. 
Then, MeV observations are crucial for verification
whether the high energy component is produced by the thermal
Comptonization in a hot flow; however, the present data quality in this
energy range is too poor for most objects to allow for such analysis. 

Thanks to its proximity, Centaurus A is the best studied FR I radio galaxy,
suitable for detailed testing of various accretion models through
modelling of its broadband spectrum. Its X-ray luminosity
($L_{2-10}/\ledd \approx 6 \times 10^{-5}$, see below, where
$L_{2-10}$ is the 2--10 keV luminosity) satisfies the criterion,
suggested in Wu, Yuan \& Cao (2007), for a FR I galaxy to be
dominated by a hot flow rather than a jet emission, of $L_{2-10} \ga 5\times 10^{-6} \ledd$. Indeed, many works attributed its spectral components to
emission from accretion flow (e.g.\ Evans et al.\ 2004; Whysong \&
Antonucci 2004; Meisenheimer et al.\ 2007) although they can also be
explained in terms of jet models (e.g.\ Abdo et al.\ 2010). Below we briefly 
compare predictions of our model with the observed properties 
of the X-ray emission in Cen A.

\noindent
(i) Luminosity. Over the last decade, Cen A was typically observed with the unabsorbed X-ray luminosity of $L_{2-10} \simeq 5 \times 10^{41}$ erg s$^{-1}$ (e.g.\ Evans et al.\ 2004, Markowitz et al.\ 2007). Our model with  $\dot m=0.1$ and $a=0.998$ gives $L_{2-10} =1.4 \times 10^{42}$ erg s$^{-1}$ (assuming $\theta_{\rm obs} = 50\degr$), which is approximately consistent with the average observed $L_{2-10}$ taking into account recent estimates of its black-hole mass of $\simeq 6 \times 10^7\, \msun$ (e.g.\ H\"aring-Neumayer et al.\ 2006), which is $\sim 3$ times smaller than $M$ assumed in our model.
Also for $\dot m=0.1$ and $a=0$ we get a roughly consistent $L_{2-10} = 10^{42}$
erg s$^{-1}$.

\noindent
(ii) Spectral index.
The time-averaged  spectral index of the X-ray emission is $\Gamma\approx 1.7$--1.8 (Beckmann et al.\ 2011, Rothschild et al.\ 2011). The model spectrum for $a=0.998$ is slightly harder, $\Gamma\approx 1.6$. The average slope for $a=0$ is roughly consistent with the observed one, but this model spectrum has pronounced scattering bumps, which have not been observed in Cen A.

\noindent
(iii) High energy cut-off. Cen A is the only FR I galaxy with data at MeV energies. Steinle et al.\ (1998) have found a spectral break in the $\Gamma\approx 1.7$--1.8 spectrum at $\sim$150 keV. Another power law spectral component was found above the break up to the MeV range. Grandi et al.\ (2003) and Beckmann et al.\ (2011) fitted the hard X-ray spectra by an e-folded power law and found e-folding energies of $\sim$500 keV, which is roughly compatible with the break energy of $\sim$150 keV. Rothschild et al.\ measured the power law spectrum up to 200 keV, within which energy range there was no indication for a break or cut-off. 

\noindent
(iv) Variability. The spectral index remains roughly unchanged while $L_{2-10}$ varies by a factor of 3 (Rothschild et al.\ 2011). In our model, changes of $L_{2-10}$ can be attributed to changes of a few different parameters, namely, the strength of the magnetic field, the fraction of accretion power directly heating electrons or the accretion rate. The results of our current study allow us to assess only effects related to changing of $\dot m$ (which case is most commonly considered as the explanation of luminosity variations), and we note that the model predicts a small change of the photon index, by $\Delta \Gamma \simeq 0.1$, corresponding to the change of $L/\ledd$ by a factor 10. This appears to be compatible with the reported spectral variability within $\Gamma\simeq 1.7$--1.8 only. 

In summary, our model can reproduce most of the properties of the X-ray emission in Cen A. However, it is inconsistent with the relatively low observed value of the break energy of $\sim 200$ keV. Furthermore, the $a=0.998$ model predicts a slightly harder spectrum than that observed. However, an additional process, which would lead to a softening of the spectrum and a decrease of the cut-off energy, has to be included in the model with $a\simeq 1$. Namely, an efficient nonthermal synchrotron emission from relativistic e$^\pm$, coming from the decay of charged pions copiously produced through proton-proton collisions in a flow around a rapidly rotating black hole, where $T_{\rm i} \ga 10^{13}$ K, would provide a strong seed photon input in addition to the thermal synchrotron emission. However, we leave a more detailed comparison of the spectrum observed from Cen A with the hot flow model spectra to a future work, in which we will take into account the relevant hadronic processes.  

\subsection{Hard spectral states}
\label{hard} 

Black-hole binaries in the hard spectral state have typical X-ray photon spectra (measured typically in the 3--10 keV range) of $\Gamma\simeq 1.5$--2 (e.g., Zdziarski \& Gierli\'nski 2004; McClintock \& Remillard 2006). In luminous spectral states, $L\ga 0.01\ledd$ or so, the high energy cut-offs in those spectra occur at energies $\simeq 50$--200 keV (e.g., Grove et al.\ 1998; Wardzi\'nski et al.\ 2002; Zdziarski \& Gierli\'nski 2004), as measured by the position of the spectral turnover in $EF_E$ spectra. 

Thus, the X-ray slope of our model spectra, $\Gamma\simeq 1.5$--1.7, is in the range of the observed values. On the other hand, the high-energy turnovers in our spectra are at $\simeq 500$ keV, which is higher than the values observed. However, our models have $L/\ledd \simeq (0.4$--$8)\times 10^{-3}$, which is lower than $L/\ledd$ at which most of the spectral cut-offs have been measured. When measured, the cut-off energy has been found to increase with the decreasing luminosity (Wardzi\'nski et al.\ 2002; Yamaoka et al.\ 2006; Miyakawa et al.\ 2008), and thus it is possible that the cut-off energy is at $\sim$500 keV in some low-$L$ hard states. 

The value of the X-ray photon index shows a negative correlation with the luminosity in luminous hard states of black-hole binaries, at $L\ga 0.01\ledd$ or so, i.e., the spectra soften with the increasing $L$. However, the correlation changes its sign at low $L$, where the spectra soften with the decreasing $L$ (e.g., Corbel et al.\ 2004; Zdziarski et al.\ 2004; Yuan et al.\ 2007; Wu \& Gu 2008; Sobolewska et al.\ 2011). 

As pointed out by Sobolewska et al.\ (2011), this indicates a change in the physical nature of the X-ray source at a luminosity of a fraction of a per cent. Those authors discuss two scenarios explaining this change. One is that the emission is dominated by an accretion flow at high $L$, and by a nonthermal jet radiation (most likely synchrotron) at low $L$. This scenario for the black-hole binary XTE J1550--564 is discussed by Russell et al.\ (2010), who found the near-infrared/optical ($\sim 10^{14.5}$ Hz) emission is tightly correlated with X-rays at low $L$, but not at high $L$. A potential problem for this model is presented by the fact that basically the same $L$-$\Gamma$ correlation, negative at low $L/\ledd$ and positive at high $L/\ledd$ is observed in AGNs. The transition is at $L/\ledd\sim 0.01$ or so, about the same as in black-hole binaries (e.g., Gu \& Cao 2009; Constantin et al.\ 2009; Younes et al.\ 2011; Veledina et al.\ 2011). The magnetic field in jets in black-hole sources is expected to scale as $B\propto M^{-1/2}$ (e.g., Heinz \& Sunyaev 2003), which (since the cyclotron frequency is $\propto B$) very strongly affects the jet emission, leading, in particular, to the strong mass dependence of the relative strength of the jet emission at a given frequency (Merloni, Heinz \& Di Matteo 2003). Thus, it appears highly unlikely that the X-ray jet emission would start to dominate the accretion-flow emission below about the same Eddington ratio in both black-hole binaries and AGNs (see also Yuan \& Cui 2005; Yuan, Yu \& Ho 2009).

The second possibility explaining the form of the $L$-$\Gamma$ correlation discussed by Sobolewska et al.\ (2011),
and independently by Veledina et al.\ (2011), is that it is due to a change of the character of the dominant seed photons for Comptonization. The overall idea follows the model for spectral changes of black-hole binaries developed by Esin et al.\ (1997, 1998), whose model did predict the changing $L$-$\Gamma$ correlation, see, e.g, fig.\ 1 in Esin et al.\ (1998). At high $L$, above the transition $L/\ledd$, the seed photons are likely to be from an optically-thick disc overlapping with the hot flow. There is a lot of observational evidence for the presence of an outer cold disc in luminous hard states (as summarized by Done, Gierli\'nski \& Kubota 2007), e.g., a strong correlation of $\Gamma$ with the relative strength of Compton reflection (Zdziarski, Lubi\'nski \& Smith 1999). Also, the inner disc radius decreases with the increasing $L$ (e.g., Done et al.\ 2007). This naturally explains the positive $L$-$\Gamma$ correlation, since the resulting increasing overlap between the cold disc and the hot flow leads to both an increase of the strength of Compton reflection and increased cooling of the Comptonizing plasma, resulting in softening of the X-ray spectra. However, at low $L/\ledd$, the outer disc may be truncated far away, and the dominant source of seed photons for Comptonization becomes synchrotron emission, as in the models studied in this work as well as in the ADAF model in general (Narayan \& Yi 1995). This model can also explain the correlation of the X-rays with the near-infrared/optical emission, which, in this scenario, would be from the ADAF flow, see Fig.\ \ref{fig:spectra_stel}. 
 
As shown in Section \ref{final}, our models spectra do harden with the increasing $L$, with $\Delta\Gamma\sim -0.1$ for $L$ increasing by a factor of several. This strongly supports the suggestion of Sobolewska et al.\ (2011) that the negative $L$-$\Gamma$ correlation seen at $L/\ledd\la 0.01$ is due to the dominance of synchrotron seed photons in that regime. A hardening of the ADAF spectra with increasing $L$, below
a few per cent of $\ledd$, was studied previously by Esin et al.\ (1997); their model seems to predict
a much stronger effect, with the increase of $\dot m$ by a factor of 10 corresponding to $\Delta\Gamma\approx -0.7$.
In this work, we have considered only two values of the accretion rate, and have not fully tested this model against observational data. We will consider this issue in detail in a paper in preparation.  

We note that Oda et al.\ (2010) propose another possible explanation for the difference between the low and high luminosity hard states. It is a transition between two equilibrium states of an optically thin, two temperature flow with a weak and a strong magnetic field, respectively. Such a change of magnetic field is outside the scope of our model.

\section{Conclusions}

We have developed a fully GR model with global Compton scattering of hot, tenuous accretion flows. The dominant source of seed photons in our model is the synchrotron process. We assume that an outer, optically-thick, accretion disc is truncated at large radii, so irradiation of the hot flow by it is negligible. We have also neglected direct viscous electron heating and outflows. We have considered the accretion rates of $\dot m = 0.1$ and 0.5, for which our model gives the bolometric luminosities of $L/\ledd \simeq (0.4$--$8)\times 10^{-3}$. 

In this range of $L/\ledd$, we find our models predict X-ray spectra hardening with increasing $L$, which is in agreement with the correlation observed at $L/\ledd\la 0.01$ in both black-hole binaries in the hard state and AGNs. This is also in agreement with the suggestion of Sobolewska et al.\ (2011) that synchrotron seed photons dominate in this range of $L/\ledd$, whereas an optically-thick accretion disc overlapping with the hot flow provide dominant source of seed photons at higher $L$. 

We have compared our models to the low-$L$ FR I radio galaxy Cen A. We have found a good agreement with the observed X-ray spectral slope, but the position of the high-energy break, observed at $\sim$200 keV, is lower than the model cut-off energy. This may be possibly explained by the hadronic processes, which would be present if the black hole in this source has a high spin. 

We have studied the impact of the black hole spin and we find that despite a very strong influence of this parameter on the dynamical properties 
of the inner part of the flow, the difference between the observed radiation from a flow surrounding a maximally rotating and a non-rotating black hole, with other parameters unchanged, is rather moderate. In particular, their luminosities differ only by a factor of 2--3 (i.e.\ smaller than a factor of $\simeq 6$ in Keplerian discs). We find this is due to a strong role of the compressive electron heating, which process only weakly depends on the black-hole spin. We note, however, that a much larger dependence on the spin value is expected in models involving direct heating of electrons, or taking into account hadronic processes. 

We stress that a proper treatment of global Compton scattering is crucial for accurate modelling of the flow structure, most importantly, for the electron 
temperature profile, and for the spectral formation in optically thin flows.
In models with $\dot m = 0.5$, we find the entire inner flow is efficiently cooled by Comptonization when the non-local nature of this process is taken into account. At $\dot m = 0.1$, global Comptonization is important, but also electron advection effects are strong at $r\ga 10$, giving rise to large temperature gradients. As compared to the results with the global Comptonization, the local model assuming a slab geometry provides in general a very poor approximation, overestimating the cooling rate by an order of magnitude in the innermost region. 

\section*{ACKNOWLEDGMENTS}

We thank the referee and F. Yuan for valuable comments. This research has been supported in part by the Polish NCN grants
N N203 582240, N N203 581240, N N203 404939 and 362/1/N-INTEGRAL/2008/09/0. FGX thanks Dr.\ Yan-Rong Li (IHEP) for discussions and sharing his GR disc
code for comparisons. FGX has been supported in part by NSFC (grants 10973003 and 10843007) and NBRPC (grants 2009CB824800 and 2009CB24901).

\label{lastpage} 
 
\end{document}